\pdfoutput=1
\documentclass{article}

\PassOptionsToPackage{numbers,sort&compress}{natbib}
\usepackage[dblblindworkshop,final]{neurips_2026}
\workshoptitle{Geometric Distributional Deep Learning (GDDL)}

\usepackage[T1]{fontenc}
\usepackage{amsmath,amssymb,amsfonts,amsthm}
\usepackage{booktabs}
\usepackage{graphicx}
\usepackage{flafter}
\usepackage{placeins}
\usepackage{microtype}
\usepackage{xcolor}
\usepackage{tikz}
\usetikzlibrary{arrows.meta,calc}
\usepackage{algorithm}
\usepackage{algpseudocode}
\usepackage{listings}
\usepackage{hyperref}
\usepackage{url}

\definecolor{certblue}{RGB}{47,88,135}
\definecolor{learnorange}{RGB}{188,98,38}
\definecolor{declgray}{RGB}{125,125,125}
\definecolor{oraclegray}{RGB}{110,110,110}
\newcommand{\R}{\mathbb{R}}
\newcommand{\N}{\mathcal{N}}
\newcommand{\E}{\mathbb{E}}
\newcommand{\KL}{\mathrm{KL}}
\newcommand{\FD}{\mathrm{D}_{\mathrm F}}
\newcommand{\F}{\mathcal{F}}
\newcommand{\eps}{\varepsilon}
\newcommand{\lcoll}{\ell_{\mathrm{coll}}}
\newcommand{\Tc}{T_{\mathrm c}}
\newcommand{\rhat}{\widehat R}

\newsavebox{\fittablebox}
\newcommand{\fittable}[1]{%
  \sbox{\fittablebox}{#1}%
  \ifdim\wd\fittablebox>\linewidth
    \resizebox{\linewidth}{!}{\usebox{\fittablebox}}%
  \else
    \usebox{\fittablebox}%
  \fi}

\newtheorem{proposition}{Proposition}
\newtheorem{corollary}[proposition]{Corollary}
\theoremstyle{remark}

\title{CONGA: Compiling Gaussian Fiber Geometry\\
for Structured Bayesian Inference}

\author{%
  V\'ictor Gallego\\
  Komorebi AI Technologies\\
  \texttt{victor.gallego@komorebi.ai}
}

\begin{document}

\maketitle

\begin{abstract}
In structured Bayesian models, a large latent block is often Gaussian
conditional on a smaller set of parameters, yet generic inference methods
relearn this geometry during warmup. We study how this structure can
be used to construct transports for posterior inference. The conditional mean
and precision determine exact affine maps, including symmetric Brenier and
triangular Knothe--Rosenblatt maps, that separate a standard normal block from
the remaining non-Gaussian core. KL and relative Fisher objectives then reduce
to objectives on this core. We introduce \textsc{conga}, a compiler that
identifies candidate Gaussian blocks and selects a temporal, spectral, or
sparse implementation according to their structure and computational cost.
Marginalization requires an identity established from the model; numerical
evidence alone permits only a target-preserving change of variables, and a
failed check leaves the model unchanged. On the
Stock--Watson inflation model, CONGA improves minimum ESS per second by
$12.5\times$ for NUTS. It also restores convergence where the raw target
fails: for WALNUTS on the same model, and for NUTS on a 6,760-dimensional
electricity-load model reduced to 34 core coordinates. Experiments on
graph-indexed fields show
how coloring and fill affect the choice of map, including cases where retaining
the original target is preferable.\\
Code released at \url{https://github.com/vicgalle/conga}.

\end{abstract}

\section{Introduction}

Normalizing flows and transport-map MCMC provide flexible ways of adapting to
posterior geometry \cite{rezende2015,hoffman2019,parno2018,papamakarios2021}.
They learn maps that account for funnels, ridges, and differences in scale,
including during HMC warmup \cite{seyboldt2026}. Learning such a map can be
difficult when the available trajectories are short or explore only a small
part of the posterior.

Our interest is in models where part of the map can be obtained analytically.
State-space models, graph fields, Gaussian-process regressions, and hierarchical
models often have a latent block $x$ that is Gaussian conditional on a smaller
nonlinear block $\eta$. We refer to the family $p(x\mid\eta)$ as a Gaussian
fiber. Its mean and precision depend on $\eta$, so a fixed whitening does not
suffice, but they give an exact whitening map for each value of $\eta$. This extends non-centering based on a
prior to a transformation based on the posterior conditional distribution
\cite{papaspiliopoulos2007,yumeng2011}.

Gaussian marginalization is well established. Its computational benefit,
however, depends on how the resulting operations are implemented. Symmetric
Brenier and triangular Knothe--Rosenblatt (KR) maps standardize the same
conditional distribution, but their costs depend on the precision graph, the
elimination order, and the operations required during inference
\cite{spantini2018,zech2022}. Choosing between these representations is the
main computational problem considered here.

We introduce \textsc{conga} (\textsc{con}ditionally \textsc{ga}ussian) to make
this choice. We first show that standardizing a Gaussian fiber separates it
from the core, reducing KL and relative Fisher objectives, as well as
compatible population-flow geometry, to the core. CONGA uses numerical probes
to identify candidate blocks and their precision structure. It marginalizes a
block only when the Gaussian identity is established from the model, then
compares temporal, spectral, and sparse implementations with inference on the
original target.

We evaluate the procedure on the Stock--Watson inflation model, a regional
electricity-load DLM, and Gaussian fields on several graph families. The
temporal models show the gains from sampling only the nonlinear core. The graph
experiments compare exact and learned transports and examine when compilation
is too expensive. Further experiments separate the effects of the backend,
sampler adaptation, and learning on the core; their protocols and results are
given in the appendices.

\begin{figure}[t]
\centering
\resizebox{\linewidth}{!}{%
\begin{tikzpicture}[
  font=\footnotesize,
  >={Stealth[length=1.9mm]},
  pbox/.style={draw=#1, fill=#1!8, rounded corners=2.5pt, line width=0.6pt,
               align=center, inner sep=3.5pt},
  gbox/.style={pbox=#1, font=\scriptsize},
  lbl/.style={font=\scriptsize},
]
\node[anchor=west, font=\footnotesize\bfseries] at (-0.25, 8.22) {(a)};
\node[lbl, font=\scriptsize\itshape, gray!40!black] at (2.70, 8.22)
  {target coordinates $(x,\eta)$};
\node[lbl, font=\scriptsize\itshape, gray!40!black] at (9.95, 8.22)
  {transported coordinates $(z,\eta)$};

\draw[->, gray!70, line width=0.5pt] (0.35,4.75) -- (5.05,4.75);
\draw[->, gray!70, line width=0.5pt] (0.35,4.75) -- (0.35,7.95);
\node[lbl, gray!30!black, anchor=north] at (2.70,4.62) {core $\eta\in\R^{k}$};
\node[lbl, gray!30!black, rotate=90] at (0.14,6.45) {block $x\in\R^{n}$};
\foreach \cx/\ytop in {1.05/5.29, 1.95/6.02, 2.85/6.51, 3.75/6.40, 4.45/5.88}{
  \draw[gray!50, dotted, line width=0.5pt] (\cx,4.80) -- (\cx,\ytop);
}
\draw[certblue!60, dashed, line width=0.6pt]
  plot[smooth, tension=0.8] coordinates
  {(0.72,5.28) (1.05,5.70) (1.95,6.50) (2.85,6.95) (3.75,6.90) (4.45,6.30) (4.75,5.92)};
\node[lbl, certblue!80!black, anchor=west, fill=white, inner sep=1pt] at (0.86,5.05) {$m(\eta)$};
\foreach \cx/\cy/\rot/\rx/\ry in {
  1.05/5.70/52/0.17/0.52,
  1.95/6.50/28/0.24/0.46,
  2.85/6.95/0/0.32/0.38,
  3.75/6.90/-32/0.22/0.50,
  4.45/6.30/-55/0.13/0.60}{
  \draw[certblue!75, fill=certblue!14, line width=0.55pt,
        rotate around={\rot:(\cx,\cy)}] (\cx,\cy) ellipse ({\rx} and {\ry});
  \draw[certblue!55, line width=0.45pt,
        rotate around={\rot:(\cx,\cy)}] (\cx,\cy) ellipse ({0.5*\rx} and {0.5*\ry});
}
\node[lbl] at (2.80,7.60)
  {fibers $p(x\mid\eta)=\N\!\big(m(\eta),\Lambda(\eta)^{-1}\big)$};

\draw[->, certblue, line width=1.0pt] (5.20,6.90) -- (7.30,6.90)
  node[midway, above=1.5pt, font=\footnotesize, certblue] {$\mathcal T_A$};
\node[lbl, certblue!90!black] at (6.10,6.55) {$z=A_\eta(x-m)$};
\node[font=\tiny, black!55] at (6.10,6.28)
  {$A_\eta=O_\eta\Lambda^{1/2}$};
\node[font=\tiny, align=center, black!60] at (6.10,5.80)
  {Brenier: $O_\eta=I$\\ KR: $A_\eta=L^\top$};

\draw[->, gray!70, line width=0.5pt] (7.60,4.75) -- (12.30,4.75);
\draw[->, gray!70, line width=0.5pt] (7.60,4.75) -- (7.60,7.95);
\node[lbl, gray!30!black, anchor=north] at (9.95,4.62) {core $\eta\in\R^{k}$};
\node[lbl, gray!30!black, rotate=90] at (7.39,6.45) {block $z\in\R^{n}$};
\foreach \cx in {8.30,9.20,10.10,11.00,11.85}{
  \draw[gray!50, dotted, line width=0.5pt] (\cx,4.80) -- (\cx,6.55);
}
\draw[gray!55, densely dashed, line width=0.5pt] (7.72,6.90) -- (12.18,6.90);
\foreach \cx in {8.30,9.20,10.10,11.00,11.85}{
  \draw[certblue!75, fill=certblue!14, line width=0.55pt] (\cx,6.90) circle (0.30);
  \draw[certblue!55, line width=0.45pt] (\cx,6.90) circle (0.16);
}
\node[lbl] at (9.95,7.60) {$z\sim\varphi_{n}$ on every fiber, independent of $\eta$};
\draw[learnorange!95!black, fill=learnorange!20, line width=0.7pt]
  plot[smooth, tension=0.75] coordinates
  {(7.80,4.77) (8.35,5.05) (8.80,5.45) (9.30,5.15) (9.85,5.10) (10.45,5.50) (11.05,5.20) (11.85,4.78)}
  -- cycle;
\node[lbl, learnorange!75!black] at (10.55,5.80) {core $p(\eta)$};

\node[lbl, align=left, certblue!90!black, anchor=west] at (12.42,6.90)
  {analytic:\\ nothing to learn};
\node[lbl, align=left, learnorange!75!black, anchor=west] at (12.42,5.20)
  {learned:\\ $S_\phi$ on $\R^{k}$};

\node[anchor=west, font=\footnotesize\bfseries] at (-0.25, 4.02) {(b)};
\node[gbox=oraclegray, text width=1.90cm] (b1) at (0.95,1.85)
  {\textsc{model or oracle}\\ algebra or $\ell,\nabla\ell$\\[0.32cm]\mbox{}};
\node[gbox=certblue, text width=2.30cm] (b2) at (3.45,1.85)
  {\textsc{fiber contract}~{\tiny\color{black!50}\S2.2}\\ $(G,H),\ m,\ \Lambda,\ \mathcal G$\\ source and checks\\[0.32cm]\mbox{}};
\node[gbox=certblue, text width=2.20cm] (b3) at (6.30,3.42)
  {\textsc{exact family}~{\tiny\color{black!50}\S2.1}\\ Brenier $\mid$ KR\\[0.32cm]\mbox{}};
\node[gbox=certblue, text width=2.85cm] (b4) at (9.22,3.42)
  {\textsc{implementation}~{\tiny\color{black!50}\S2.2}\\ \mbox{temporal $\mid$ spectral $\mid$ sparse}\\[0.32cm]\mbox{}};
\node[gbox=certblue, text width=2.10cm] (b5) at (12.02,3.42)
  {\textsc{workload}~{\tiny\color{black!50}\S2.2}\\ build, score, draws\\[0.32cm]\mbox{}};
\node[gbox=certblue, densely dashed, text width=3.35cm] (b6) at (8.60,1.85)
  {\textsc{change of variables}~{\tiny\color{black!50}\S2.2}\\ \mbox{exact transformed target}\\[0.32cm]\mbox{}};
\node[gbox=declgray, text width=2.10cm, text=black!65] (b7) at (8.60,0.28)
  {\textsc{unchanged}\\ original target\\[0.32cm]\mbox{}};
\node[gbox=learnorange, text width=1.90cm] (b8) at (13.30,1.85)
  {\textsc{inference}~{\tiny\color{black!50}\S3}\\ \mbox{HMC $+$ core $S_\phi$}\\[0.32cm]\mbox{}};
\begin{scope}[shift={([yshift=0.31cm]b1.south)}]   %
  \draw[-{Stealth[length=1mm]}, black!50, line width=0.45pt] (-0.21,0.09) -- (-0.05,-0.06);
  \draw[-{Stealth[length=1mm]}, black!50, line width=0.45pt] (0.21,0.09) -- (0.05,-0.06);
  \fill[black!40] (-0.25,0.12) circle (0.05) (0.25,0.12) circle (0.05) (0,-0.10) circle (0.05);
\end{scope}
\begin{scope}[shift={([yshift=0.31cm]b2.south)}]   %
  \draw[certblue!75, line width=0.6pt] (0,0.01) circle (0.15);
  \draw[certblue!85!black, line width=0.7pt] (-0.07,0.01) -- (-0.02,-0.05) -- (0.09,0.10);
\end{scope}
\begin{scope}[shift={([yshift=0.31cm]b3.south)}]   %
  \draw[certblue!75, fill=certblue!14, line width=0.5pt,
        rotate around={35:(-0.40,0)}] (-0.40,0) ellipse (0.09 and 0.20);
  \draw[-{Stealth[length=1mm]}, certblue!70, line width=0.5pt] (-0.15,0) -- (0.13,0);
  \draw[certblue!75, fill=certblue!14, line width=0.5pt] (0.33,0) circle (0.12);
\end{scope}
\begin{scope}[shift={([yshift=0.31cm]b4.south)}]   %
  \draw[certblue!60, line width=0.5pt]
    (-0.42,-0.06) -- (-0.14,-0.06) -- (0.14,-0.06) -- (0.42,-0.06)
    (-0.14,-0.06) -- (0.0,0.14) -- (0.14,-0.06);
  \foreach \p in {(-0.42,-0.06),(-0.14,-0.06),(0.14,-0.06),(0.42,-0.06),(0.0,0.14)}
    \fill[certblue!75] \p circle (0.045);
\end{scope}
\begin{scope}[shift={([yshift=0.29cm]b5.south)}]   %
  \foreach \x/\h in {-0.21/0.12, -0.05/0.26, 0.11/0.18}
    \draw[certblue!70, fill=certblue!30, line width=0.4pt] (\x,-0.11) rectangle ++(0.11,\h);
\end{scope}
\begin{scope}[shift={([yshift=0.31cm]b6.south)}]   %
  \draw[certblue!70, line width=0.5pt] plot[smooth, tension=0.7]
    coordinates {(-0.52,-0.09) (-0.36,0.10) (-0.20,-0.09)};
  \node[font=\tiny, certblue!80!black, inner sep=0pt] at (0,0) {$=$};
  \draw[certblue!70, line width=0.5pt] plot[smooth, tension=0.7]
    coordinates {(0.20,-0.09) (0.36,0.10) (0.52,-0.09)};
\end{scope}
\begin{scope}[shift={([yshift=0.31cm]b7.south)}]   %
  \draw[-{Stealth[length=1mm]}, black!45, line width=0.5pt]
    (0.11,0.04) arc[start angle=20, end angle=330, radius=0.12];
\end{scope}
\begin{scope}[shift={([yshift=0.31cm]b8.south)}]   %
  \draw[learnorange!90!black, line width=0.55pt] plot[smooth, tension=0.8]
    coordinates {(-0.38,-0.06) (-0.27,0.09) (-0.16,-0.03) (-0.03,0.11)
                 (0.09,0.00) (0.21,0.10) (0.33,-0.04)};
\end{scope}
\draw[gray!60] (b1.east) -- (b2.west);
\draw[certblue!70] (b2.east) -- (5.00,1.85);
\fill[certblue!70] (5.00,1.85) circle (1.3pt);
\draw[->, certblue!80] (5.00,1.85) |- (b3.west);
\node[font=\tiny, certblue!80!black, anchor=east] at (4.90,2.95) {exact};
\draw[->, certblue!80] (b3.east) -- (b4.west);
\draw[->, certblue!80] (b4.east) -- (b5.west);
\draw[->, certblue!80, densely dashed] (5.00,1.85) -- (b6.west);
\node[font=\tiny, certblue!80!black, anchor=south, inner sep=1.5pt] at (6.10,1.85) {numerical};
\draw[->, declgray] (5.00,1.85) |- (b7.west);
\node[font=\tiny, black!55, anchor=east] at (4.90,0.85) {invalid / uneconomic};
\draw[->, certblue!80] (b5.east) -| (b8.north);
\draw[->, certblue!80, densely dashed] (b6.east) -- (b8.west);
\draw[->, declgray] (b7.east) -| (b8.south);

\end{tikzpicture}%
}
\caption{The CONGA workflow. (a)~A Gaussian fiber determines the exact
family in Eq.~\eqref{eq:transport-family}; its Brenier and KR members share
factorization (Proposition~\ref{prop:factorization}) and core-only learning
(Corollary~\ref{cor:objectives}). (b)~For an exact contract, CONGA chooses an
implementation from the graph and the workload. Numerical candidates retain
the target through Eq.~\eqref{eq:failsafe}; invalid or expensive candidates
leave it unchanged.}
\label{fig:framework}
\end{figure}
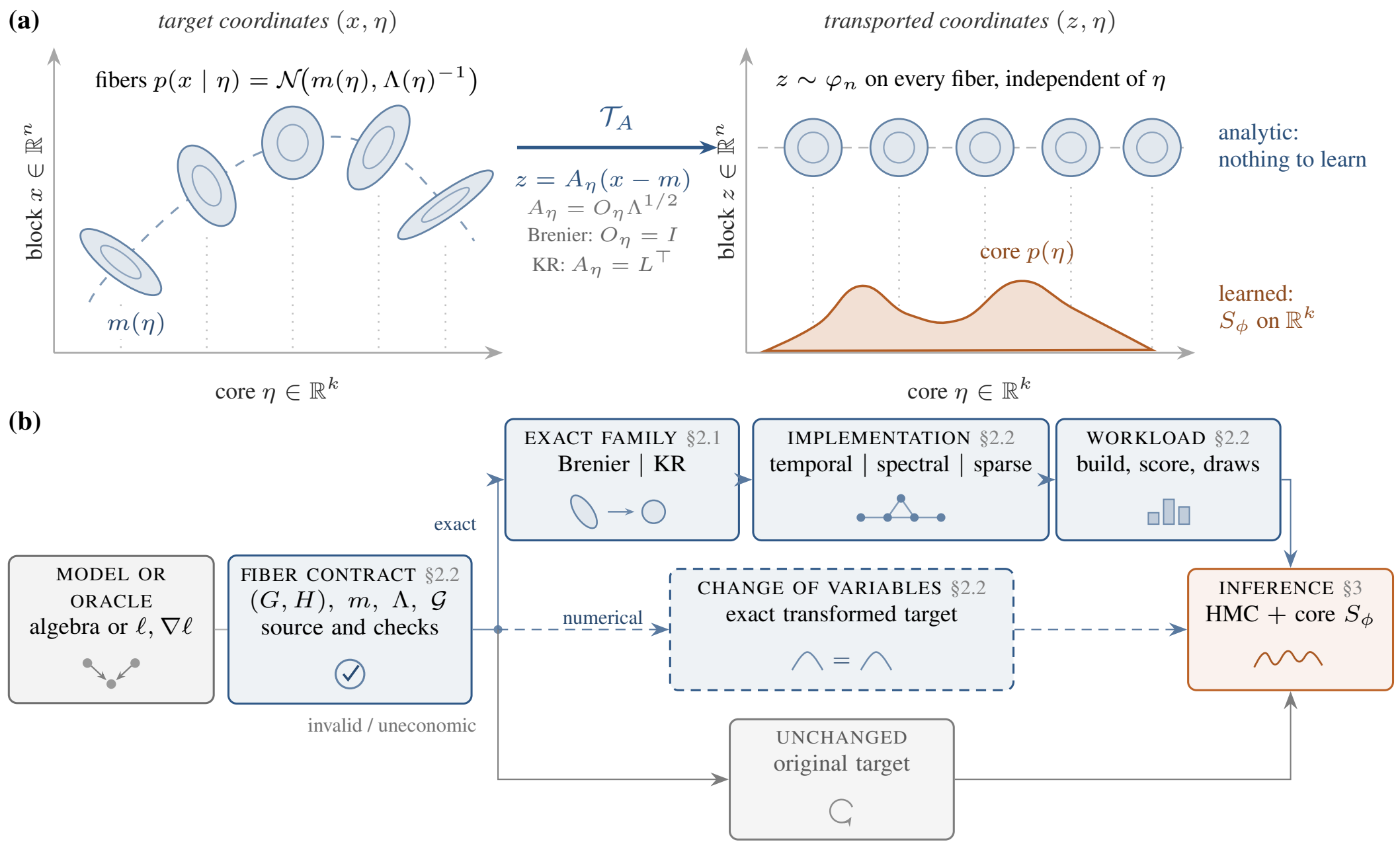

\section{The CONGA framework}
\label{sec:framework}

CONGA has two parts: the exact transports determined by a Gaussian fiber,
and the compiler that selects among them.

\subsection{Gaussian fiber transports}
\label{sec:geometry}

\subsubsection{Gaussian fibers factorize exactly}

Let $p(x,\eta)\propto\exp\ell(x,\eta)$ on $\R^n\times\R^k$. Suppose that, for
every $\eta$ in the relevant domain,
\begin{equation}
  p(x\mid\eta)=\N\!\left(x;m(\eta),\Lambda(\eta)^{-1}\right),
  \qquad \Lambda(\eta)=L(\eta)L(\eta)^\top\succ0,
  \label{eq:conditional}
\end{equation}
and write $\Lambda^{1/2}$ for the symmetric positive-definite square root. Every
affine standardizer of this conditional law has the form
\begin{equation}
 A_\eta=O_\eta\Lambda(\eta)^{1/2},\qquad O_\eta^\top O_\eta=I,
 \quad
 \mathcal T_A(x,\eta)=\bigl(A_\eta(x-m(\eta)),\eta\bigr).
 \label{eq:transport-family}
\end{equation}
Conversely, $A_\eta\Lambda(\eta)^{-1}A_\eta^\top=I$ implies this form.
The orthogonal factor therefore describes the freedom in choosing the
standardized coordinates.

The following conditional-Gaussian identity provides the basis for the
compiler.

\begin{proposition}[Exact fiber factorization]
\label{prop:factorization}
If Eq.~\eqref{eq:conditional} holds globally and $\eta\mapsto O_\eta$ is
measurable, then every map in Eq.~\eqref{eq:transport-family} satisfies
\begin{equation}
  (\mathcal T_A)_\#p(z,\eta)=\varphi_n(z)\,p(\eta),
  \label{eq:factorization}
\end{equation}
where $\varphi_n=\N(0,I_n)$. The transported block is standard normal and
independent of the core.
\end{proposition}

\begin{proof}
Conditioned on $\eta$, Eq.~\eqref{eq:transport-family} maps the stated Gaussian
to $\N(0,I_n)$. This conditional law is the same for every $\eta$, hence $z$ is
independent of $\eta$ and the joint density is
$\varphi_n(z)p(\eta)$.
\end{proof}

\subsubsection{Brenier and KR implementations}

The symmetric choice $O_\eta=I$ gives $A_\eta=\Lambda^{1/2}$, the Brenier map
from $\N(m,\Lambda^{-1})$ to $\N(0,I)$, and minimizes quadratic displacement
within each fiber \cite{peyre2019}. This is a fiberwise statement among maps
that preserve $\eta$, not a global optimal-transport claim for the joint law.
If instead $\Lambda=LL^\top$ for a lower-triangular Cholesky factor, then
$A_\eta=L^\top$ gives the triangular KR representative
\begin{equation}
  \Tc(x,\eta)=\bigl(L(\eta)^{\top}(x-m(\eta)),\eta\bigr)=(z,\eta),
  \quad
  \Tc^{-1}(z,\eta)=\bigl(m(\eta)+L(\eta)^{-\top}z,\eta\bigr).
  \label{eq:transport}
\end{equation}
This map is block triangular even when $m$ and $L$ depend on $\eta$, with
$\log|\det J_{\Tc^{-1}}|=-\log|L(\eta)|$. Its coordinate order determines
sparse fill and dependency depth \cite{spantini2018,zech2022}. Both maps satisfy
Proposition~\ref{prop:factorization}, although their implementations can be
quite different.

\subsubsection{Learning on the core}

Let $S_\phi:\R^k\to\R^k$ be any learned residual flow and
$F_\phi=(\mathrm{id}_z\otimes S_\phi)\circ\mathcal T_A$. Let $\FD$ denote
relative Fisher divergence to a standard normal.

\begin{corollary}[Learning reduces to the core]
\label{cor:objectives}
Under Proposition~\ref{prop:factorization}, for every member of the transport
family,
\begin{align}
 \KL\bigl((F_\phi)_\#p\,\|\,\varphi_{n+k}\bigr)
 &=\KL\bigl((S_\phi)_\#p(\eta)\,\|\,\varphi_k\bigr),\\
 \FD\bigl((F_\phi)_\#p\,\|\,\varphi_{n+k}\bigr)
 &=\FD\bigl((S_\phi)_\#p(\eta)\,\|\,\varphi_k\bigr).
 \label{eq:objective-reduction}
\end{align}
Both objectives depend only on the transformed core distribution.
\end{corollary}

At $S_\phi=\mathrm{id}$, the remaining objective is
$\FD(p(\eta)\|\varphi_k)$. Transforming the standardized fiber cannot improve
it. For the state-space example below, this reduces the learned map from
$8+2T$ dimensions to eight.

The factorization also applies to compatible population dynamics. Define
$\iota(q)=\varphi_n\otimes q$ and the direct-sum kernel
\begin{equation}
 \mathbf K_\oplus((z,\eta),(z',\eta'))
 =\operatorname{diag}\!\left(k_z(z,z')I_n,
 k_\eta(\eta,\eta')I_k\right).
 \label{eq:direct-sum-kernel}
\end{equation}

\begin{proposition}[Fiber-reduced gradient flows]
\label{prop:gradient-flow}
For regular $q_0,q_1$ and the Euclidean product cost, $\iota$ is isometric for
quadratic Wasserstein, spherical Fisher--Rao and mass-preserving WFR distances.
Under Eq.~\eqref{eq:direct-sum-kernel}, $D\iota$ also preserves Stein tangent
norms, so the intrinsic geometry restricted to $\mathcal M_n$ is the core
geometry. If $\Pi=\varphi_n\otimes\pi$, all four KL gradient vectors at
$\iota(q)$ are tangent lifts of their core counterparts. Thus unique flow
semigroups commute with $\iota$. Under an exact $\mathcal T_A$, the same holds
for $p(dx\mid\eta)q(d\eta)$ in the pullback geometry.
\end{proposition}

The proof, together with the corresponding finite-particle qualification, is
given in Appendix~\ref{app:gradient-flow}. A generic joint kernel need not
preserve this set of measures, so the result does not apply to an arbitrary
finite particle cloud.

\subsection{Compiling fiber contracts}
\label{sec:pass}

A \emph{fiber contract} records the block/core partition, the conditional
moments $m(\eta)$ and $\Lambda(\eta)$, the structural graph of $\Lambda$, and
how these quantities were established. The compiler uses this information to
choose a valid implementation of the Gaussian identity.

\subsubsection{Discovery and validation}
\label{sec:certificate}

We first identify coordinates with a jointly quadratic log density using
secants of the score. Directional differences recover the precision pattern,
after which colored probes recover its nonzero entries. When the precision
is positive definite, we obtain candidate conditional moments and test them
at fresh points through
\begin{equation}
 r(x,\eta)=\ell(x,\eta)-\lcoll(\eta)
 -\log\N\!\left(x;m(\eta),\Lambda(\eta)^{-1}\right),
 \label{eq:residual}
\end{equation}
where $\lcoll$ is the Gaussian collapsed log density. These checks give a
\emph{numerical certificate}; they do not prove that $r\equiv0$. CONGA removes
the residual only when model algebra or symbolic analysis establishes the
identity. It can then use the collapsed log density and score
\begin{equation}
 \tilde\ell(z,\eta)=\lcoll(\eta)-\tfrac12\|z\|^2-\tfrac n2\log(2\pi),
 \qquad
 \nabla\tilde\ell=(-z,\nabla_\eta\lcoll),
 \label{eq:compiled-score}
\end{equation}
up to coordinate order. If only a numerical certificate is available, the
candidate map is used as an ordinary change of variables,
\begin{equation}
 \tilde\ell(u)=\ell(T^{-1}u)+\log|\det J_{T^{-1}}(u)|,
 \label{eq:failsafe}
\end{equation}
which remains correct even if the proposed structure is wrong. When validation
fails, the model is left unchanged. Appendix~\ref{app:proof} gives an
approximation bound, and Appendices~\ref{app:interface}--\ref{app:discovery}
describe the complete pass.

\subsubsection{Choosing an implementation}

The graph determines how much work each implementation needs: coloring
controls precision recovery, fill controls triangular factorization, and
bandwidth controls banded storage. We compare marginalization,
preconditioning, and inference on the unchanged target. For backend $a$,
workload $W$, and $R$ uses of the compiled object, we estimate the cost as
\[
 \widehat C_a(W,R)=B_a+R\widehat n_a(W)t_a,
\]
where $B_a$ is construction time, $t_a$ is score latency, and $\widehat n_a$
is a pilot estimate of the number of score evaluations. Invalid programs are
discarded, while the original target is always available. An exact rewrite may
therefore be rejected simply because it costs more.

\paragraph{Population check.}\label{sec:census}
We also ran the discovery procedure once on 46 targets, without
model-specific configuration. CONGA proposed fibers for 38 targets, found none
for seven, and rejected one random-probe candidate at the held-out pattern
check. All accepted numerical residuals were below $2.3\times10^{-9}$ in
relative terms, but none was treated as an exact contract. On the pilots
target, sampling through the discovered change of variables has no
divergences, against a median of 450 for the raw sampler. The full census is
reported in Appendix~\ref{app:census}.

\subsubsection{Graph-dependent implementations}

The maps in Eq.~\eqref{eq:transport-family} give the same factorization.
To choose an implementation, we use the graph structure and the intended
workload, as summarized in Table~\ref{tab:lowerings}.

\begin{table}[ht]
 \centering
 \footnotesize
 \setlength{\tabcolsep}{4pt}
 \caption{Implementations available for different fiber structures.}
 \label{tab:lowerings}
 \fittable{%
 \begin{tabular}{@{}lll@{}}
  \toprule
  structure & implementation & selection criteria \\
  \midrule
  temporal path & sequential or balanced elimination & dependency depth; build reuse \\
  regular graph pencil & fast-transform Brenier & shared spectrum; fast basis \\
  general sparse graph & triangular KR & distance-two coloring; fill \\
  invalid or expensive candidate & original target & validation; workload cost \\
  \bottomrule
 \end{tabular}}
\end{table}

On a temporal path, we choose between sequential Kalman elimination and a
balanced associative scan. For regular graph pencils we cache a spectral
Brenier map; for general sparse graphs we use a triangular KR map with a
fill-reducing order. The choice depends on the cost of construction, score
evaluation, and reconstruction. Figure~\ref{fig:lowerings} and
Appendix~\ref{app:gmrf} describe these graph programs.
Appendix~\ref{app:spectral} examines the spectral choice and
Appendix~\ref{app:scaling} the temporal one, where balanced elimination
overtakes the sequential scan for scores from $T=400$.

For sparse graphs, recovery requires a coloring of the squared graph $G^2$,
whereas execution follows the filled Cholesky graph. For example, a star has no fill yet requires $|V|$ colors, while an
expander needs few probes but produces substantial fill. We therefore check
both quantities before running an expensive pilot.

\section{Experiments}
\label{sec:decision}
\label{sec:core}

We evaluate CONGA on three models with different Gaussian fiber structures.
Within each sampling comparison, we match seeds and compute diagnostics in
the same reconstructed coordinates. Reported efficiencies include construction
and reconstruction, and a method passes only when every seed has
rank-$\rhat\le1.05$. Full protocols and per-seed
results are given in Appendices~\ref{app:gmrf}, \ref{app:stock-watson}, and
\ref{app:dlm-load}.

\subsection{Stock--Watson inflation}
\label{sec:stock-main}

For quarterly inflation $y_{1:T}$, the Stock--Watson model
\cite{stockwatson2007} has persistent and transient log-volatility paths
$z$ and $x$, and a latent level $\tau$:
\begin{align}
 z_t\mid z_{t-1},\sigma&\sim\N(z_{t-1},\sigma^2), &
 x_t\mid x_{t-1},\sigma&\sim\N(x_{t-1},\sigma^2),\nonumber\\
 \tau_t\mid\tau_{t-1},z_{t-1}&\sim\N(\tau_{t-1},e^{z_{t-1}}), &
 y_t\mid\tau_t,x_t&\sim\N(\tau_t,e^{x_t}),
 \label{eq:stock-main}
\end{align}
with $\sigma^{-2}\sim\operatorname{Gamma}(5,0.5)$. At $T=252$, the raw
innovation target has 756 coordinates. Conditional on $(z,x)$, the centered
$\tau_{1:T}$ path is Gaussian with tridiagonal precision, so CONGA integrates
252 coordinates and samples the 504-dimensional volatility core.

Across three seeds and eight chains, marginalizing $\tau$ reduces median NUTS
time from 1,821 to 406 seconds and WALNUTS time from 682 to 289 seconds. For NUTS,
where both parameterizations pass, minimum bulk ESS per second improves by
$12.5\times$. Raw WALNUTS fails the convergence gate in every seed, so we
report no efficiency ratio for it; its CONGA counterpart passes. The difference between NUTS and WALNUTS remains after compilation:
marginalizing the Gaussian block and adapting the local integrator address
different parts of the posterior geometry.

\subsection{Electricity-load DLM}
\label{sec:load-main}

Let $\mathbf s_t$ collect regional trends, harmonic seasonals, dynamic
temperature responses, holiday effects, and a national factor. The model is a
conditionally linear-Gaussian state-space system,
\begin{align}
 \mathbf s_t\mid\mathbf s_{t-1},\eta
   &\sim\N\!\left(F_t(\eta)\mathbf s_{t-1},Q_t(\eta)\right),\nonumber\\
 \mathbf y_t\mid\mathbf s_t,\eta
   &\sim\N\!\left(H_t(\eta)\mathbf s_t,
     \operatorname{diag}_{r}\{e^{x_{d(t)}+c_r}\}\right).
 \label{eq:load-main}
\end{align}
For example, regional trend innovations have covariance
$e^{z_{d(t)}}(Q_{\mathcal G}+\kappa I)^{-1}$ on the transmission graph, while
the national factor obeys
$f_t\mid f_{t-1},\eta\sim\N(\varphi_f f_{t-1},e^{z^f_{d(t)}})$.
The core $\eta$ contains the daily volatility paths, $\kappa$, dynamic roots,
factor loadings, observation offsets, and innovation scales; conditional on it,
the entire trajectory $\mathbf s_{1:T}$ is Gaussian.

At the matched scale $R=3,T=168$, the fiber has 6,726 coordinates and the core
has 34. Its precision is block-tridiagonal but neither commuting nor Toeplitz,
so CONGA uses a sequential Kalman program rather than a spectral map. Raw NUTS
has a minimum ESS of about five and fails the rank-$\rhat$ gate in all three
seeds. CONGA reaches a minimum ESS of 1,161 and a worst $\rhat=1.006$ in
comparable time. Since the raw chains do not converge, this is a difference
between failing and passing rather than a speedup. The same
program evaluates one value-and-gradient of a 4.5-million-coordinate extension
in 214 seconds.

\begin{table}[t]
 \centering
 \footnotesize
 \setlength{\tabcolsep}{3.4pt}
 \caption{End-to-end temporal-model results. Sampling times and minimum bulk ESS
 are three-seed medians; efficiency ratios in the text additionally include
 reconstruction. Rank-$\rhat$ is the worst seed and divergences are summed.}
 \label{tab:end-to-end-main}
 \fittable{%
 \begin{tabular}{@{}llrrrrr@{}}
  \toprule
  model & arm & sampled dim. & seconds & min. ESS & worst $\rhat$ & div.\\
  \midrule
  Stock--Watson & NUTS raw & 756 & 1,821 & 437 & 1.020 & 64\\
   & NUTS + CONGA & 504 & 406 & 1,207 & 1.004 & 4\\
   & WALNUTS raw & 756 & 682 & 37 & 1.168 & --\\
   & WALNUTS + CONGA & 504 & 289 & 416 & 1.022 & --\\
  \addlinespace
  load DLM & NUTS raw & 6,760 & 631 & 5.3 & 2.121 & 30\\
   & NUTS + CONGA & 34 & 566 & 1,161 & 1.006 & 3\\
  \bottomrule
 \end{tabular}}
\end{table}

\subsection{Graph-indexed fields}
\label{sec:gmrf}

For a graph $G$ with Laplacian $Q_G$, we use the properized intrinsic-CAR
smoother
\begin{align}
 u&\sim\N(-1,0.7^2), & v&\sim\N(-1.5,0.7^2),\nonumber\\
 x\mid u&\sim\N\!\left(0,e^{2u}(Q_G+\kappa I)^{-1}\right), &
 y\mid x,v&\sim\N(x,e^{2v}I),
 \label{eq:gmrf-model}\\[1mm]
 \Lambda(u,v)&=e^{-2u}(Q_G+\kappa I)+e^{-2v}I,
 \label{eq:gmrf-precision}
\end{align}
with $\kappa=0.2$. The $n$-dimensional field $x$ is the fiber and $(u,v)$ is
the core. Paths and square lattices share a fast cosine basis and admit a
Brenier map; general graphs use colored precision recovery and sparse KR.

\begin{figure}[t]
 \centering
 \includegraphics[width=\linewidth]{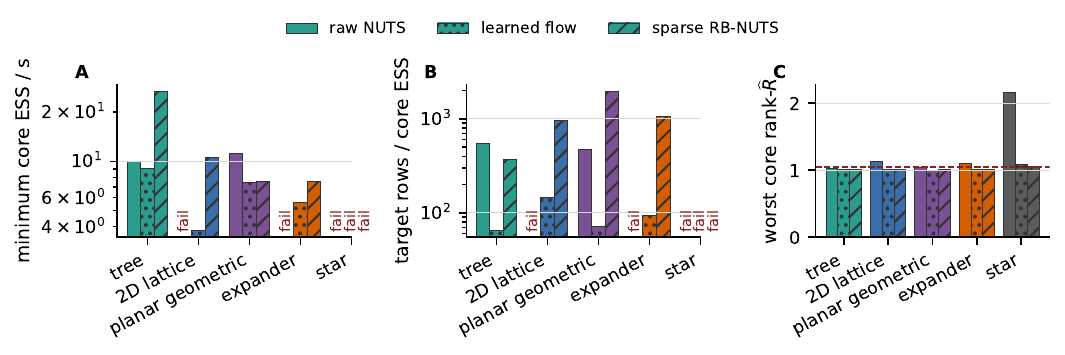}
 \caption{Matched inference on graph fields at $n=256$ (three-seed medians).
 Minimum core ESS/s and target-equivalent rows per ESS are shown only when every
 seed passes rank-$\rhat\le1.05$; time includes discovery and reconstruction.
 Complete values and protocols are in Appendix~\ref{app:gmrf}.}
 \label{fig:gmrf-nuts}
\end{figure}

Figure~\ref{fig:gmrf-nuts} compares raw NUTS, a learned flow, and sparse
Rao--Blackwellized NUTS on trees, lattices, planar graphs, expanders, and stars.
Sparse collapse passes on the tree, lattice, planar graph, and expander. It has
the highest median ESS/s on the tree, lattice, and expander; on the planar graph
it is comparable to the learned flow. The flow uses fewer target rows but
incurs the cost of fitting and evaluating the map. No method passes on the
star. Since its complete square graph requires 256 probe colors, CONGA can
reject this exact collapse before sampling.

\subsection{Additional experiments}
\label{sec:experiments}
\label{sec:scope}
\label{sec:dbn}

Three smaller studies separate effects that are combined in the end-to-end
comparisons.

\paragraph{Backend choice.}
With the collapsed posterior and sampler fixed at
$n=1{,}024$, fast-transform
Brenier reduces seconds per minimum core ESS by $7.7\times$ on a path and
$38.1\times$ on a lattice relative to sparse KR. The maps are statistically
equivalent, so this difference comes from their implementations
(Appendix~\ref{app:spectral}).

\paragraph{Learning on the core.}
Let
$s_t\mid s_{t-1},\eta\sim\N(As_{t-1},Q)$ be a two-state DLM with
Gaussian observations and an eight-dimensional parameter core. At $T=100$,
CONGA removes the 200-dimensional state fiber before fitting the same residual
flow. The eight-dimensional flow reaches median held-out Fisher divergence
$0.52$, versus $5.45\times10^4$ for the 208-dimensional flow, with 19,776 rather
than 279,376 parameters. In the matching sampling comparison, the compiled
transport is $10.8$--$15.0\times$ faster than a generic adapter for the same
map, through both compilation and metric adaptation, while the learned arms
fail the convergence criterion at every length
(Appendices~\ref{app:scaling}--\ref{app:attribution}). Core-size sweeps and
diagnostics are in
Appendices~\ref{app:core}--\ref{app:graphcore}.

\paragraph{Keeping the original target.}
Two cases show when CONGA does so. The star graph of Section~\ref{sec:gmrf} is
conditionally Gaussian, but recovering its precision needs one probe color
per node, which is too expensive. As a second control, we give the two-state
DLM above Student-$t$ sensor noise, so that the
Gaussian identity fails. Applying the Gaussian candidate through
Eq.~\eqref{eq:failsafe} remains calibrated, whereas removing its residual
biases posterior means (Appendix~\ref{app:misspec}). These examples distinguish the validity of a
transformation from its computational benefit.

\section{Related work and limitations}
\label{sec:related}\label{sec:limitations}

\paragraph{Transport maps.}
Conditional independence induces sparsity in triangular KR maps, which allows
a learned transport to scale with Markov structure \cite{spantini2018,zech2022}.
The Knothe rearrangement is also an anisotropic limit of quadratic-cost
transport \cite{carlier2010}. Here we consider an exactly solvable Gaussian
fiber and use the nonuniqueness of its whitening map to choose between symmetric
Brenier and ordered triangular representations. Transport-map MCMC and
normalizing flows learn the map instead
\cite{parno2018,rezende2015,hoffman2019,papamakarios2021}, including during HMC
warmup \cite{seyboldt2026}. CONGA first removes the part that is known exactly.
Appendix~\ref{app:gradient-flow} relates the same factorization to SVGD and other
measure-valued flows \cite{liu2017,duncan2023,chizat2018}.

\paragraph{Compilation.}
Gaussian marginalization and Rao--Blackwellization are classical, as are
conditional reconstruction, sparse GMRF algebra, and Kalman filtering
\cite{casella1996,kalman1960,rueheld2005,rue2009}. Probabilistic-programming
systems exploit conjugacy and elimination from syntax, traces, or intermediate
representations \cite{obermeyer2019,murray2018,lai2023,hoffman2018,
gorinova2020}. CONGA starts from oracle-based discovery, checks the proposed
structure, and selects among several exact implementations according to the
workload.

\paragraph{Limitations.}
Finite probes cannot prove a global identity, so marginalization still requires
an algebraic or symbolic contract. The current implementation covers temporal
and graph templates rather than arbitrary kernels. It does not infer commuting
matrix pencils, and it reconstructs joint temporal draws sequentially.
Non-Gaussian or discrete fibers would require different maps. Finally, the
experiments use three seeds and a limited model census. They assess the procedure on these
models; the choice of representation remains dependent on the target and
workload.

\section{Conclusions}

We have introduced CONGA, a compiler for conditionally Gaussian blocks in
structured Bayesian models. Starting from an exact fiber identity, it chooses a
transport using the precision graph and the intended workload. The Gaussian
block can then be removed from sampling and transport learning. When the
identity is supported only numerically, the proposed map is used as a
target-preserving reparameterization instead. In our experiments, the choice of
exact implementation changes runtime substantially, and fitting a learned map
only on the nonlinear core is much more stable than fitting it on the full
state. These results suggest that model-derived geometry is worth using before
estimating a more general transport from samples.

\paragraph{AI-Use Disclosure.}
AI coding assistants (Claude Code) were used to help implement and run the
experiments, to write analysis and plotting scripts, and to draft and edit
parts of the manuscript. The authors designed the method and the experiments,
checked the results, and take full responsibility for the content of the
paper.

\clearpage
\appendix

\section{Theory, compiler, and discovery details}
\label{app:theory}

\subsection{Fiber-reduced gradient flows}
\label{app:gradient-flow}

This appendix proves Proposition~\ref{prop:gradient-flow}. We work with smooth,
positive densities having sufficient decay for the integrations by parts below;
the dynamic-metric statements extend by their usual relaxation. Write
$Z=\R^n$, $H=\R^k$, $X=Z\times H$, $\gamma=\varphi_n$, and
$E\psi(z,\eta)=\psi(\eta)$. For a core tangent $\sigma$,
$D\iota_q[\sigma]=\gamma\otimes\sigma$. The two identities used throughout
are
\begin{equation}
 \frac{d\iota(q)}{d\iota(\pi)}(z,\eta)=\frac{dq}{d\pi}(\eta),
 \qquad \int_X E\psi\,d\iota(q)=\int_H\psi\,dq.
 \label{eq:cylindrical-identities}
\end{equation}

\paragraph{Dynamic-metric isometries.}
For $W_2$, coupling the same $z\sim\gamma$ and an optimal core pair gives
$W_2(\iota(q_0),\iota(q_1))\le W_2(q_0,q_1)$. Projection onto $H$ is
1-Lipschitz and gives the reverse inequality. For Fisher--Rao and WFR, consider
the continuity--reaction equation
\begin{equation}
 \partial_t\rho_t+\nabla_z\cdot(\rho_t v_t^Z)
 +\nabla_\eta\cdot(\rho_t v_t^H)=\rho_t a_t
 \label{eq:wfr-continuity}
\end{equation}
with action
\begin{equation}
 \mathcal A_X=\int_0^1\!\int_X
 \{c_T(\lVert v_t^Z\rVert^2+\lVert v_t^H\rVert^2)
 +c_R|a_t|^2\}\,d\rho_t\,dt,
 \label{eq:wfr-action}
\end{equation}
where the fixed positive weights encode the chosen WFR convention. A core path
$(q_t,u_t,b_t)$ lifts as
$\rho_t=\gamma\otimes q_t$, $v_t=(0,u_t)$, and $a_t=b_t$ with identical
action. Conversely, project any admissible ambient path and define conditional
averages
\begin{equation}
 q_t=(\operatorname{pr}_H)_\#\rho_t,
 \quad u_t(\eta)=\E_{\rho_t}[v_t^H\mid\eta],
 \quad b_t(\eta)=\E_{\rho_t}[a_t\mid\eta].
 \label{eq:wfr-projection}
\end{equation}
They satisfy
$\partial_tq_t+\nabla_\eta\cdot(q_tu_t)=q_tb_t$, and conditional Jensen
plus omission of the nonnegative $z$-transport term gives
$\mathcal A_H(q,u,b)\le\mathcal A_X(\rho,v,a)$. Taking infima proves equality.
The same argument with transport or reaction suppressed proves the pure
Fisher--Rao case (and recovers $W_2$). If paths are restricted to probabilities,
$\int a_t\,d\rho_t=0$ projects to $\int b_t\,dq_t=0$; without this restriction
the proof also covers the standard unbalanced FR and WFR metrics.

\paragraph{Compatible Stein tangent geometry.}
The RKHS of Eq.~\eqref{eq:direct-sum-kernel} is the orthogonal sum
$\mathcal H_z^n\oplus\mathcal H_\eta^k$. Define the Stein tangent norm by
\begin{equation}
 \lVert\sigma\rVert_{q,\mathrm{St}}^2
 =\inf_{u\in\mathcal H_\eta^k}
 \{\lVert u\rVert_{\mathcal H_\eta^k}^2:
 \sigma=-\nabla_\eta\cdot(qu)\}.
 \label{eq:stein-tangent}
\end{equation}
If $u$ generates $\sigma$, then $(0,u)$ generates
$D\iota_q[\sigma]$ with the same norm. Conversely, suppose
$v=(v_Z(z),v_H(\eta))$ in the direct-sum RKHS generates this lifted tangent.
After division by $\gamma(z)$, its continuity equation is
\begin{equation}
 -q(\eta)\,\frac{\nabla_z\cdot(\gamma v_Z)}{\gamma}(z)
 -\nabla_\eta\cdot(qv_H)(\eta)=\sigma(\eta).
 \label{eq:stein-separation}
\end{equation}
The first quotient must be constant in $z$; its integral against $\gamma$ is
zero by the boundary condition. Hence it vanishes, $v_H$ is admissible in
Eq.~\eqref{eq:stein-tangent}, and direct-sum orthogonality supplies the reverse
norm inequality. Thus $D\iota_q$ is isometric on tangent spaces and the
intrinsic Stein geometry restricted to $\mathcal M_n$ is exactly the core
geometry. We do not claim that every nonunique ambient Stein geodesic
remains in the leaf.

\paragraph{Gradient-flow commutation.}
Let $\mathsf K^G_\rho$ be the Onsager operator for geometry $G$. For
Wasserstein, spherical Fisher--Rao and WFR it is, up to the fixed metric
weights, respectively
\begin{align}
 \mathsf K^W_\rho\xi&=-\nabla\cdot(\rho\nabla\xi),\nonumber\\
 \mathsf K^{\mathrm{FR}}_\rho\xi
 &=\rho\left(\xi-\int\xi\,d\rho\right),\qquad
 \mathsf K^{\mathrm{WFR}}_\rho
 =c_T'\mathsf K^W_\rho+c_R'\mathsf K^{\mathrm{FR}}_\rho.
 \label{eq:onsager-operators}
\end{align}
For Stein geometry, use the kernel mobility induced by
Eq.~\eqref{eq:direct-sum-kernel}. Direct substitution, and integration of the
fixed fiber where needed, gives the single intertwining identity
\begin{equation}
 \mathsf K^G_{\iota(q)}E\psi
 =D\iota_q[\mathsf K^{G,H}_q\psi],
 \qquad G\in\{W,\mathrm{FR},\mathrm{WFR},\mathrm{St}\}.
 \label{eq:onsager-intertwining}
\end{equation}
By Eq.~\eqref{eq:cylindrical-identities},
\begin{equation}
 \left.\frac{\delta}{\delta\rho}\KL(\rho\|\Pi)
 \right|_{\rho=\iota(q)}
 =E\!\left(\log\frac{q}{\pi}+1\right).
 \label{eq:kl-first-variation}
\end{equation}
Equations~\eqref{eq:onsager-intertwining}--\eqref{eq:kl-first-variation}
show that the ambient KL gradient vector is $D\iota_q$ applied to its core
counterpart. Therefore every core solution lifts to an ambient solution, and
uniqueness makes the corresponding semigroups commute.

Finally, for an exact CONGA map $\mathcal T_A$, set
\begin{equation}
 \iota_A(q)=(\mathcal T_A^{-1})_\#(\gamma\otimes q)
 =p(dx\mid\eta)q(d\eta),
 \qquad
 d_G^A(P_0,P_1)=d_G((\mathcal T_A)_\#P_0,(\mathcal T_A)_\#P_1).
 \label{eq:pullback-geometry}
\end{equation}
The compiled result applied after pushforward proves the pullback claim. It does
not assert isometry for ordinary Euclidean $W_2$ in $(x,\eta)$, because
$\mathcal T_A^{-1}$ generally depends on $\eta$. Population SVGD has zero fiber
velocity by Gaussian Stein identity, but this cancellation is not exact for an
empirical cloud; a finite-particle implementation should evolve core particles
and reconstruct the fiber. Likewise, ordinary WFR means Wasserstein transport
plus Fisher--Rao reaction. Combining a kernel-mobility Langevin sampler such as
stochastic-gradient Langevin dynamics with repulsion (SGLD+R) with that reaction
is instead a generalized Stein--Fisher--Rao Onsager sum, although
Eq.~\eqref{eq:onsager-intertwining} still reduces each compatible component to
the core.

\begin{figure}[t]
 \centering
 \includegraphics[width=\linewidth]{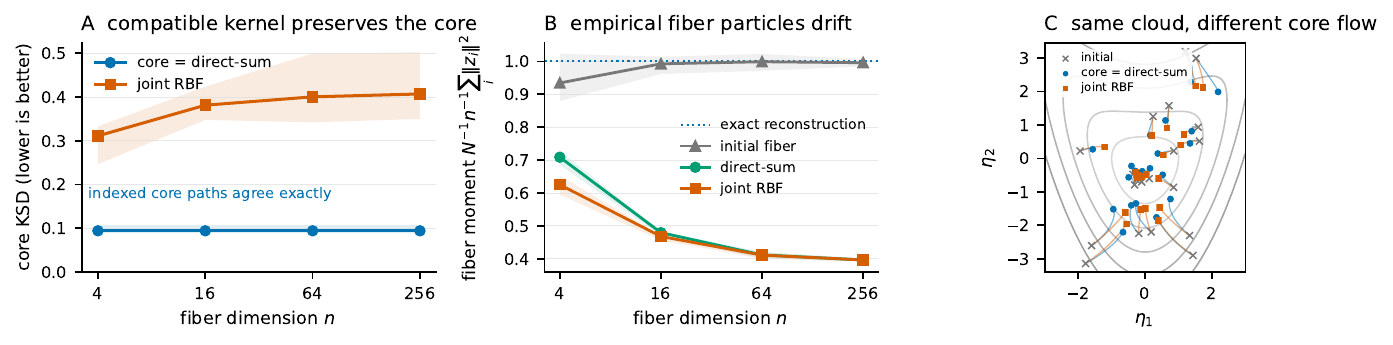}
 \caption{Finite-particle consequence and boundary of
 Proposition~\ref{prop:gradient-flow} on
 $\varphi_n\otimes\pi_{\mathrm{banana}}$ ($N=64$, 300 steps, five seeds; medians
 and ranges). (A) Core-only and direct-sum paths coincide; a joint RBF couples
 core updates to fiber distances. (B) Finite Gaussian clouds are not stationary,
 whereas core-only reconstruction preserves the exact second moment. (C) At
 $n=256$, compatible paths follow the banana geometry; joint-RBF paths are
 altered.}
 \label{fig:svgd-fiber}
\end{figure}

\paragraph{Finite-particle SVGD protocol.}
\label{app:svgd-fiber}
Figure~\ref{fig:svgd-fiber} uses the two-dimensional banana core
\begin{equation}
 \eta_1\sim\N(0,1),\qquad
 \eta_2\mid\eta_1\sim\N\!\left(0.35(\eta_1^2-1),1\right),
 \label{eq:banana-core}
\end{equation}
whose score is available analytically, and an independent
$z\sim\N(0,I_n)$ fiber for $n\in\{4,16,64,256\}$. Each of five seeds initializes
the same $N=64$ indexed core particles from $1.6\,\N(0,I_2)$ and the same
ambient particles from
$\N(0,I_n)$. We compare core-only SVGD, ambient SVGD with
Eq.~\eqref{eq:direct-sum-kernel}, and ambient SVGD with one scalar RBF on
$(z,\eta)$. In all three arms, the RBF blocks use the same functional form and
median heuristic,
\begin{equation}
 k(u,u')=\exp(-\lVert u-u'\rVert^2/h),\qquad
 h=\operatorname{median}_{i\ne j}\lVert u_i-u_j\rVert^2/\log(N+1),
 \label{eq:svgd-bandwidth}
\end{equation}
recomputed at every step. The direct-sum core bandwidth depends only on
$\eta$, as required by the theorem; the joint bandwidth uses full-coordinate
distances. We apply 300 explicit Euler steps of size $0.1$ to the standard
$N^{-1}$-normalized SVGD velocity. Each final core cloud is scored by the same
V-statistic KSD, using the paired core-only bandwidth. Bands show the full seed
range. Panel C was fixed before execution as seed 0 at the largest fiber,
$n=256$. It overlays analytic negative-log-density contours with the initial
and final core positions and trajectories of twenty particle indices selected by
equal spacing through the indexed cloud.

The direct-sum and core-only updates are evaluated independently and compared
after every step; their maximum coordinate discrepancy is zero in float64 in
all 20 cells. Halving the step size to $0.05$ and doubling the steps at fixed
integration time changes the reported fiber second moments by at most $0.001$
in the checked $n\in\{4,64,256\}$ cells.

\subsection{Robustness of an approximate fiber contract}
\label{app:proof}

The main text uses exact contracts for residual-erasing compilation and the
target-preserving density in Eq.~\eqref{eq:failsafe} for numerical candidates.
For completeness, a uniform error bound also controls the distributional effect
of an approximate identity.

\begin{proposition}[Uniform approximate identity]
\label{prop:robustness}
Let $q_0(z,\eta)=\varphi_n(z)p(\eta)$ be the ideal factorized target. If the
residual of Eq.~\eqref{eq:residual} satisfies $|r(x,\eta)|\le\eps$ globally, then
the actual pushforward $\tilde p$ satisfies
\begin{equation}
 \KL(\tilde p\|q_0)\le2\eps,
 \qquad \|\tilde p-q_0\|_{\mathrm{TV}}\le\sqrt{\eps}.
 \label{eq:robustness}
\end{equation}
Moreover, an independence proposal from the candidate conditional Gaussian,
corrected against the original target, has pointwise acceptance at least
$e^{-2\eps}$.
\end{proposition}

\begin{proof}
Let $q_\eta=\N(m(\eta),\Lambda(\eta)^{-1})$ and
$C(\eta)=\E_{q_\eta}e^{r(X,\eta)}$. Integration over $x$ shows that, for
$\tilde r=r\circ\Tc^{-1}$,
\[
  \frac{\tilde p(z,\eta)}{q_0(z,\eta)}
  =\frac{e^{\tilde r(z,\eta)}}{C(\eta)}.
\]
Both factors lie in $[e^{-\eps},e^{\eps}]$, so the log ratio is in
$[-2\eps,2\eps]$; expectation and Pinsker give the two bounds. At fixed $\eta$,
$C(\eta)$ cancels from the corrected proposal ratio, giving acceptance
$\min\{1,e^{r(x',\eta)-r(x,\eta)}\}\ge e^{-2\eps}$.
\end{proof}

\subsection{Oracle discovery and structural report}
\label{app:interface}

Algorithm~\ref{alg:pass} gives the complete pass. Listing~\ref{lst:api} shows
the released calls for the numerical path and Listing~\ref{lst:api-full} the
structural reports they return.

\begin{algorithm}[ht]
\small
\caption{CONGA: from a fiber contract to a workload-selected program.}
\label{alg:pass}
\begin{algorithmic}[1]
\Require exact fiber contract or value-and-gradient oracle; backends $\mathcal B$;
workload $W$
\If{no contract is supplied}
  \State discover $(G,H)$; recover $m,\Lambda,\mathcal G$; validate $r$
  \State form a \emph{numerical} contract with its validation record
\EndIf
\If{validation fails} \Return unchanged target \EndIf
\State price coloring, fill, bandwidth; $\mathcal A\gets\{\text{unchanged}\}$
\If{authorization is exact}
  \State expose $A_\eta=O_\eta\Lambda^{1/2}$; add available graph lowerings
  \Comment{temporal, spectral, sparse KR}
\Else
  \State add the target-preserving reparameterization of Eq.~\eqref{eq:failsafe}
\EndIf
\State measure build/score/draw costs; pilot score counts and diagnostics
\State \Return lowest-cost diagnostically valid program in $\mathcal A$
\end{algorithmic}
\end{algorithm}

\begin{lstlisting}[float=ht,caption={Released numerical path. The explicit calls
expose a candidate contract and its costs; \texttt{fit\_structure\_metric}
performs the same discovery internally and returns a whitening metric, which
preserves the target as in Eq.~\eqref{eq:failsafe}. Exact residual-erasing
backends additionally require the model contract of Algorithm~\ref{alg:pass}.},
label={lst:api}]
from conga import (
    GaussianBlock, discover_gaussian_block, fit_structure_metric)
found = discover_gaussian_block(target)
fiber = GaussianBlock(
    target, found.g_idx, probe_etas=found.probe_etas)
fiber.d_g, fiber.d_h, fiber.n_colors, fiber.bandwidth
metric = fit_structure_metric(logp_and_grad, dim)
\end{lstlisting}

\begin{algorithm}[ht]
\caption{Oracle discovery subroutine used in Algorithm~\ref{alg:pass}.}
\label{alg:discovery}
\begin{algorithmic}[1]
\Require oracle $\theta\mapsto(\ell,\nabla\ell)$ on $\R^{D}$; relative tolerance
 $\mathrm{tol}=10^{-8}$; $n_{\mathrm{base}}$ probe points
\Ensure candidate block/core frontier $\{G_k,H_k\}$, moments
 $m_k(\cdot),L_k(\cdot)$, invariants $(c_k,\mathrm{fill}_k,b_k)$, numerical
 validation record
\State \textbf{axis probes:} secant slopes of $\nabla\ell$ at two step sizes;
 relative disagreement $>\mathrm{tol}$ marks a conflict edge
 \Comment{$n_{\mathrm{base}}(1+4D)$ gradient rows}
\State \textbf{peel:} drop self-conflicted coordinates, then remove
 maximum-degree vertices until the conflict graph is edge-free
\State \textbf{screen:} mixed second differences along random joint directions
 in $\operatorname{span}(G)$; peel the argmax offender until below tolerance
\If{$G=\emptyset$} \Return invalid contract: sampler unchanged \EndIf
\State \textbf{pattern:} directional score differences give the sparsity of
 $\Lambda$; distance-two coloring gives $c$, reverse Cuthill--McKee the bandwidth
 $b$ and the dense or banded backend; enumerate maximum-degree hub peels and
 retain the nondominated $(d_g,d_h,c,\mathrm{fill},b)$ prefixes
\State \textbf{moments and certificate:} colored assembly of $\Lambda(\eta)$,
 Cholesky $\Lambda=LL^{\top}$, $m=\Lambda^{-1}g_0$; verify symmetry,
 colored-versus-column assembly, positive definiteness, and the held-out residual
 $r$ of Eq.~\eqref{eq:residual} at fresh $(x,\eta)$, rejecting otherwise
\State \textbf{return:} numerical fiber contract and structural price report;
 exact authorization remains external
\end{algorithmic}
\end{algorithm}

\begin{lstlisting}[float=ht,caption={Abridged structural reports corresponding
to Listing~\ref{lst:api}. Discovery reads the oracle only, so both records have
numerical authorization; exact marginalization in the compiled experiments uses
a separate model-level contract.},label={lst:api-full}]
from conga import discover_gaussian_block, GaussianBlock, fit_structure_metric

found = discover_gaussian_block(target)      # oracle in: values and gradients
block = GaussianBlock(target, found.g_idx, probe_etas=found.probe_etas)
block.d_g, block.d_h, block.n_colors, block.bandwidth      # invariants out
metric = fit_structure_metric(logp_and_grad, dim)   # whitening metric for a sampler

# report for pilots
D = 18   d_g = 15   d_h = 3
probe colors = 14   bandwidth = 12   banded backend = False
held-out identity residual |r| = 7.11e-14 nats  (3.5e-16 relative)
authorization = numerical (oracle evidence only)
programs = {precondition, unchanged}  # exact contract may add marginalize

# report for diamonds
D = 26   d_g = 24   d_h = 2   probe colors = 24   bandwidth = 23
held-out identity residual |r| = 2.18e-11 nats (1.7e-16 relative)
authorization = numerical (oracle evidence only)
programs = {precondition, unchanged}  # exact contract may add marginalize
\end{lstlisting}

\FloatBarrier

\subsection{The discovery front-end}
\label{app:discovery}

The discovery procedure takes a value-and-gradient oracle for $\ell$ on
$\R^D$ and returns a candidate Gaussian block $G$ and a core $H$. We use the
same defaults throughout. The following describes the probes, selection rule,
and validation checks in Algorithm~\ref{alg:discovery}.

\paragraph{What is being tested.}
$\ell$ is quadratic in the coordinates of $G$, for fixed $\eta=\theta_H$, exactly
when every third derivative $\partial^3\ell/\partial\theta_i\partial\theta_j
\partial\theta_k$ with $i,j,k\in G$ vanishes. Rather than form third derivatives,
we use the equivalent statement that the secant slopes of the score along
directions in $\operatorname{span}(G)$ are independent of step size: for a
quadratic they are exact at any $\delta$. The pass therefore needs first-order
gradient rows only, batched, with no Hessian, no higher-order automatic
differentiation, and no access to model syntax.

\paragraph{Stage 1: axis probes and the conflict graph.}
We draw $n_{\mathrm{base}}$ base points ($3$ in the census) as
$0.5\,\N(0,I_D)$ in the unconstrained space, redrawing until $\ell$ and
$\nabla\ell$ are finite (at most $50\,n_{\mathrm{base}}$ attempts, after which
the procedure returns an error). At each base point $\theta$ and for each
step pair $(\delta_1,\delta_2)\in\{(0.3,0.9),(0.7,2.1)\}$ we form the axis
secants
\begin{equation}
 h^{(\delta)}_{ik}
 =\frac{\partial_i\ell(\theta+\delta e_k)-\partial_i\ell(\theta)}{\delta},
 \qquad
 R_{ik}=\max\frac{\bigl|h^{(\delta_2)}_{ik}-h^{(\delta_1)}_{ik}\bigr|}
 {\max\bigl\{|h^{(\delta_1)}_{ik}|,\,|h^{(\delta_2)}_{ik}|,\,\text{floor}_k\bigr\}},
 \label{eq:axis-residual}
\end{equation}
the maximum running over base points and step pairs. The floor
$\max\{10^{-3}|h_{kk}|,10^{-9}\max_{ik}|h_{ik}|\}$ keeps structurally zero
entries from producing spurious relative residuals. Symmetrizing
$R>\mathrm{tol}$ gives a conflict graph in which an edge $i\!-\!k$ witnesses the
hyperedge $\{i,k,k\}$; its diagonal marks coordinates along whose own axis
$\ell$ is not quadratic, as can occur for log scales, lengthscales, and
non-Gaussian priors. Axes with non-finite gradients are also excluded. Stage 1 costs
$n_{\mathrm{base}}(1+4D)$ gradient rows, streamed in chunks so that neither
$\mathrm{eye}(D)$ nor the full probe batch is materialized.

\paragraph{Stage 2: greedy peel.}
We drop the self-conflicted coordinates and then repeatedly remove a
maximum-degree vertex of the remaining conflict graph until it is edge-free.
Finding a maximum independent set is NP-hard, so this rule does not guarantee
the largest block. In the state-space model, the transition coefficients have
degrees growing with $T$ and are removed first, leaving the $2T$ states used
in Section~\ref{sec:dbn}.

\paragraph{Stage 3: joint quadraticity screening.}
Purely trilinear hyperedges $\{i,j,k\}$ with distinct indices are invisible to
axis probes. We therefore evaluate the mixed second difference
\begin{equation}
 \Delta(u,v)=\nabla\ell(\theta+u+v)-\nabla\ell(\theta+u)
 -\nabla\ell(\theta+v)+\nabla\ell(\theta),
 \label{eq:joint-residual}
\end{equation}
with $u,v\sim0.5\,\N(0,I)$ supported on the current candidate set. $\Delta$
vanishes identically on $\operatorname{span}(G)$ if and only if $\ell$ is jointly
quadratic there, at any step size. We use four direction pairs per base point,
normalize $\Delta$ by the observed first differences, and peel the argmax
offending coordinate until the residual falls below tolerance, at most $32$ rounds.
Testing finitely many pairs provides numerical evidence of quadraticity. Each
pair requires four gradient rows; in the census, the screening cost was small
relative to the axis probes.

\paragraph{Tolerance, and the boundary it draws.}
The single tolerance is relative, $\mathrm{tol}=10^{-8}$, roughly five orders of
magnitude above float64 secant roundoff. It is deliberately strict enough to
exclude densities that are only softly non-quadratic: a wide Student-t prior is
non-quadratic at the $10^{-6}$ level and is rejected, keeping the emitted collapse
exact rather than approximate. On 49 of the 50 census models the decision is not
marginal: the largest normalized residual among \emph{kept} coordinates is
$2.4\times10^{-9}$ (median $2.7\times10^{-13}$), the smallest residual on a
\emph{flagged} pair is $9.3\times10^{-8}$ (median $0.56$), and the final joint
residual never exceeds $2.3\times10^{-12}$---a bimodal separation of at least a
factor of 39 and usually many orders of magnitude. The exception is mcycle\_gp,
which is the extreme on both sides at once: kept residual $4.2\times10^{-9}$
against smallest flagged $1.0\times10^{-8}$, straddling tol by less than a factor
of three, with joint residual $2.38\times10^{-10}$. It is also the one model whose
block is unstable across probe seeds and the one whose candidate the held-out
pattern check rejects (Appendix~\ref{app:census}). The margin statistics therefore
locate the hard case without being told about it, and it is the only model for which
the boundary discussion applies.

\paragraph{Do the constants matter?}
Using one fixed set of constants everywhere establishes consistency, not
correctness, so we rerun the whole census under seven alternatives---three other
step-pair sets, including a single pair rather than two; base scales $0.2$ and
$1.0$; and tolerances $10^{-6}$ and $10^{-10}$. The sweep costs $14.0$ seconds of
discovery in total. The step and scale variants leave the split at $42$ validated,
$7$ declined and $1$ rejected in every case, with no model changing verdict and only
mcycle\_gp's emitted block moving at all (between $47$ and $50$ coordinates); this
is what quadraticity predicts, since the secant slope of a quadratic is exact at any
step. The tolerance is the one load-bearing constant, and it fails safely in both
directions. At $10^{-6}$ the diamonds block grows from $24$ to $25$ coordinates by
admitting a softly non-quadratic one, and the held-out pattern check rejects it, so
the split becomes $41/7/2$: loosening the screen cannot admit an approximate collapse through the
downstream checks; it converts a validated model into a rejected one. At $10^{-10}$ the split is again $42/7/1$, but six blocks shrink
(rats $62\to33$, three kidiq variants $4\to3$ or $2$, earnings $2\to1$, mcycle
$49\to40$); tightening only surrenders coordinates to the core and never invalidates
a collapse. The split therefore depends on the targets rather than the probe schedule.
Of the constants we vary, only the exactness tolerance changes it.

\paragraph{Restarts and probe placement.}
The census uses seeds $0,1,2$, with seed 0 as the primary result. Random probes
(mode A) give the same index set across seeds on 49 of 50 models and agree
with probes at reference draws (mode B) on 48 of 50. Mode B assesses the
effect of placement when posterior information is already available.

For an implementable alternative, we place both discovery and validation
probes at draws from a mean-field variational fit, using 2,000 steps and eight
Monte Carlo samples per step without model-specific tuning. This reproduces
mode B's block on all 46 targets where both are defined: 45 PosteriorDB
posteriors with shipped references and the DBN. In particular, all three fit
seeds recover the 40-coordinate mcycle\_gp block, compared with the rejected
49-coordinate candidate from random probes. Its retained and flagged
residuals then separate to $2.2\times10^{-11}$ and $1.1\times10^{-5}$.
The four supplementary real-data targets are outside this comparison.

\paragraph{From candidate block to map.}
Given $G$, the sparsity pattern of $\Lambda(\eta)$ is detected by three further
probes, retaining entries above $10^{-9}$ relative to the largest entry and
taking the union over probes; a reverse Cuthill--McKee bandwidth then selects the
dense or banded backend (banded when $d_g\ge512$ and $4(\text{bandwidth}+1)\le
d_g$, a measured CPU crossover). Construction verifies, at a held-out probe, that
the recovered Hessian is symmetric and that colored assembly reproduces
column-by-column assembly, both to $10^{-8}$ relative, and that the banded and
dense collapsed densities agree to $10^{-9}$; positive definiteness and the
identity residual of Eq.~\eqref{eq:residual} are then checked at posterior-plausible
cores. Quadraticity does not imply definiteness, and these are separate checks:
a jointly quadratic block whose recovered precision is indefinite somewhere has no
conditional Gaussian there, and the pass declines.

\paragraph{Cost-aware refinement.}
A large block can be expensive to recover if its precision graph contains
hubs. We successively remove maximum-degree vertices and retain the
nondominated tuples $(d_g,d_h,c,\mathrm{fill},b)$. Each sub-block uses a
principal submatrix of the detected pattern, so this step requires no further
model evaluations. The workload determines which tuple to use. For radon,
moving two global means into the core changes the $172/3$ split with 147
colors to a $170/5$ split with two colors. For pilots, the short diagnostic
run favors retaining all 15 block coordinates despite the 14 colors.

\paragraph{Implementation limits.}
The output is a numerical certificate. Exact marginalization in the DBN and
GMRF experiments additionally uses model algebra. The procedure recovers
indices, precision structure, and conditional moments, but does not synthesize
an arbitrary collapsed kernel: the sequential and balanced Gaussian-message
templates are supplied by hand. Automatically translating recovered temporal
paths and elimination graphs into such templates remains future work.

\FloatBarrier
\section{Additional experiments}
\label{app:additional}

\subsection{Graph-indexed fields: protocol and complete results}
\label{app:gmrf}

\begin{figure}[t]
\centering
\resizebox{\linewidth}{!}{%
\begin{tikzpicture}[
  font=\footnotesize,
  >={Stealth[length=1.6mm]},
  lbl/.style={font=\scriptsize},
  tin/.style={font=\tiny, black!60},
  emit/.style={draw=certblue!60, fill=certblue!6, rounded corners=2pt,
               inner sep=2.5pt, font=\tiny, text=certblue!60!black},
]
\draw[gray!35, line width=0.4pt] (4.70,0.15) -- (4.70,4.30);
\draw[gray!35, line width=0.4pt] (9.60,0.15) -- (9.60,4.30);

\begin{scope}
\node[lbl] at (2.25,4.45) {\textbf{(a)} temporal path: choose a schedule};
\draw[certblue!55, line width=0.6pt] (0.45,3.85) -- (3.75,3.85);
\foreach \k in {0,...,7}
  \fill[certblue!80] ({0.45+0.4714*\k},3.85) circle (0.055);
\node[tin] at (0.45,3.60) {$s_1$};
\node[tin] at (3.75,3.60) {$s_T$};
\node[tin] at (2.10,3.38) {two exact elimination orders};
\foreach \k in {0,...,7} \fill[certblue!70] ({0.30+0.22*\k},1.15) circle (0.04);
\draw[certblue!70, line width=0.55pt] (0.30,1.15) -- (1.84,3.11);
\foreach \k in {1,...,7}
  \draw[certblue!70, line width=0.55pt]
    ({0.30+0.22*\k},1.15) -- ({0.30+0.22*\k},{1.15+0.28*\k});
\foreach \k in {1,...,7}
  \fill[certblue!95!black] ({0.30+0.22*\k},{1.15+0.28*\k}) circle (0.045);
\foreach \k in {0,...,7} \fill[certblue!70] ({2.55+0.22*\k},1.15) circle (0.04);
\foreach \j in {0,...,3}{
  \draw[certblue!70, line width=0.55pt]
    ({2.55+0.44*\j},1.15) -- ({2.66+0.44*\j},1.43) -- ({2.77+0.44*\j},1.15);
  \fill[certblue!95!black] ({2.66+0.44*\j},1.43) circle (0.045);
}
\draw[certblue!70, line width=0.55pt] (2.66,1.43) -- (2.88,1.71) -- (3.10,1.43);
\draw[certblue!70, line width=0.55pt] (3.54,1.43) -- (3.76,1.71) -- (3.98,1.43);
\fill[certblue!95!black] (2.88,1.71) circle (0.045) (3.76,1.71) circle (0.045);
\draw[certblue!70, line width=0.55pt] (2.88,1.71) -- (3.32,1.99) -- (3.76,1.71);
\fill[certblue!95!black] (3.32,1.99) circle (0.045);
\node[tin, align=center] at (1.07,0.78) {sequential (Kalman)\\ depth $O(T)$};
\node[tin, align=center] at (3.32,0.78) {balanced scan\\ depth $O(\log T)$};
\node[emit] at (2.25,0.28) {emit: same total work, schedule priced per workload};
\end{scope}

\begin{scope}[shift={(4.9,0)}]
\node[lbl] at (2.25,4.45) {\textbf{(b)} regular pencil: choose Brenier};
\foreach \i in {0,...,3}{
  \draw[certblue!50, line width=0.5pt] (0.50,{2.60+0.40*\i}) -- (1.70,{2.60+0.40*\i});
  \draw[certblue!50, line width=0.5pt] ({0.50+0.40*\i},2.60) -- ({0.50+0.40*\i},3.80);
}
\foreach \i in {0,...,3} \foreach \j in {0,...,3}
  \fill[certblue!80] ({0.50+0.40*\i},{2.60+0.40*\j}) circle (0.045);
\node[tin] at (1.10,2.33) {path / lattice $Q_G$};
\draw[->, certblue!80, line width=0.7pt] (1.95,3.20) -- (2.40,3.20);
\node[tin, certblue!80!black] at (2.17,3.46) {cache $U$};
\draw[certblue!85, line width=0.6pt]
  plot[domain=0:1, samples=48, smooth] ({2.55+1.65*\x},{3.62+0.14*cos(180*\x)});
\draw[certblue!60, line width=0.6pt]
  plot[domain=0:1, samples=48, smooth] ({2.55+1.65*\x},{3.20+0.14*cos(360*\x)});
\draw[certblue!40, line width=0.6pt]
  plot[domain=0:1, samples=48, smooth] ({2.55+1.65*\x},{2.78+0.14*cos(540*\x)});
\node[tin] at (3.38,2.33) {fast basis: DCT / Kronecker};
\draw[gray!35, line width=0.4pt] (0.72,1.06) rectangle (1.50,1.84);
\foreach \i in {0,...,5}
  \fill[certblue!90] ({0.735+0.13*\i},{1.73-0.13*\i}) rectangle ++(0.10,0.10);
\foreach \i in {0,...,4}{
  \fill[certblue!45] ({0.735+0.13*\i},{1.60-0.13*\i}) rectangle ++(0.10,0.10);
  \fill[certblue!45] ({0.865+0.13*\i},{1.73-0.13*\i}) rectangle ++(0.10,0.10);
}
\foreach \i in {0,...,2}{
  \fill[certblue!25] ({0.735+0.13*\i},{1.34-0.13*\i}) rectangle ++(0.10,0.10);
  \fill[certblue!25] ({1.125+0.13*\i},{1.73-0.13*\i}) rectangle ++(0.10,0.10);
}
\draw[->, certblue!80, line width=0.7pt] (1.72,1.45) -- (2.42,1.45);
\node[tin, certblue!80!black] at (2.07,1.70) {$U^\top(\,\cdot\,)\,U$};
\draw[gray!35, line width=0.4pt] (2.62,1.06) rectangle (3.40,1.84);
\foreach \i in {0,...,5}
  \fill[certblue!90] ({2.635+0.13*\i},{1.73-0.13*\i}) rectangle ++(0.10,0.10);
\node[tin] at (3.01,0.86) {diagonal for every $\eta$};
\node[emit] at (2.25,0.28) {emit: spectral map $+$ collapsed score, transform speed};
\end{scope}

\begin{scope}[shift={(9.8,0)}]
\node[lbl] at (2.25,4.45) {\textbf{(c)} general sparse graph: choose KR};
\foreach \a in {0,60,...,300}{
  \draw[certblue!55, line width=0.5pt] (1.05,3.30) -- ++(\a:0.55);
  \fill[certblue!80] ($(1.05,3.30)+(\a:0.55)$) circle (0.04);
}
\fill[certblue!95!black] (1.05,3.30) circle (0.055);
\node[tin] at (1.05,2.52) {star};
\foreach \i in {0,...,7}{
  \draw[certblue!45, line width=0.45pt]
    ($(3.35,3.30)+(45*\i:0.55)$) -- ($(3.35,3.30)+(45*\i+45:0.55)$);
  \draw[certblue!45, line width=0.45pt]
    ($(3.35,3.30)+(45*\i:0.55)$) -- ($(3.35,3.30)+(45*\i+135:0.55)$);
}
\foreach \i in {0,...,7} \fill[certblue!80] ($(3.35,3.30)+(45*\i:0.55)$) circle (0.04);
\node[tin] at (3.35,2.52) {expander};
\draw[black!40, line width=0.4pt] (0.62,1.45) -- (1.48,1.45);
\draw[certblue!70, fill=certblue!45, line width=0.4pt] (0.70,1.45) rectangle ++(0.22,0.85);
\draw[learnorange!80, fill=learnorange!45, line width=0.4pt] (1.16,1.45) rectangle ++(0.22,0.07);
\draw[black!40, line width=0.4pt] (2.92,1.45) -- (3.78,1.45);
\draw[certblue!70, fill=certblue!45, line width=0.4pt] (3.00,1.45) rectangle ++(0.22,0.22);
\draw[learnorange!80, fill=learnorange!45, line width=0.4pt] (3.46,1.45) rectangle ++(0.22,0.85);
\draw[certblue!70, fill=certblue!45, line width=0.4pt] (0.50,0.76) rectangle ++(0.13,0.13);
\node[tin, anchor=west] at (0.67,0.825) {probe: colors $c(G^2)$};
\draw[learnorange!80, fill=learnorange!45, line width=0.4pt] (2.78,0.76) rectangle ++(0.13,0.13);
\node[tin, anchor=west] at (2.95,0.825) {run: fill $\operatorname{nnz}(L)$};
\node[emit] at (2.25,0.28) {emit: KR in fill order --- or unchanged if both high};
\end{scope}
\end{tikzpicture}%
}
\caption{One contract, three graph lowerings (Table~\ref{tab:lowerings}).
(a)~Path schedules trade $O(T)$ for $O(\log T)$ dependency depth. (b)~A cached
fast basis diagonalizes a regular precision pencil for every $\eta$. (c)~General
graphs separate probing cost $c(G^2)$ from Cholesky fill; high costs return the
unchanged target.}
\label{fig:lowerings}
\end{figure}
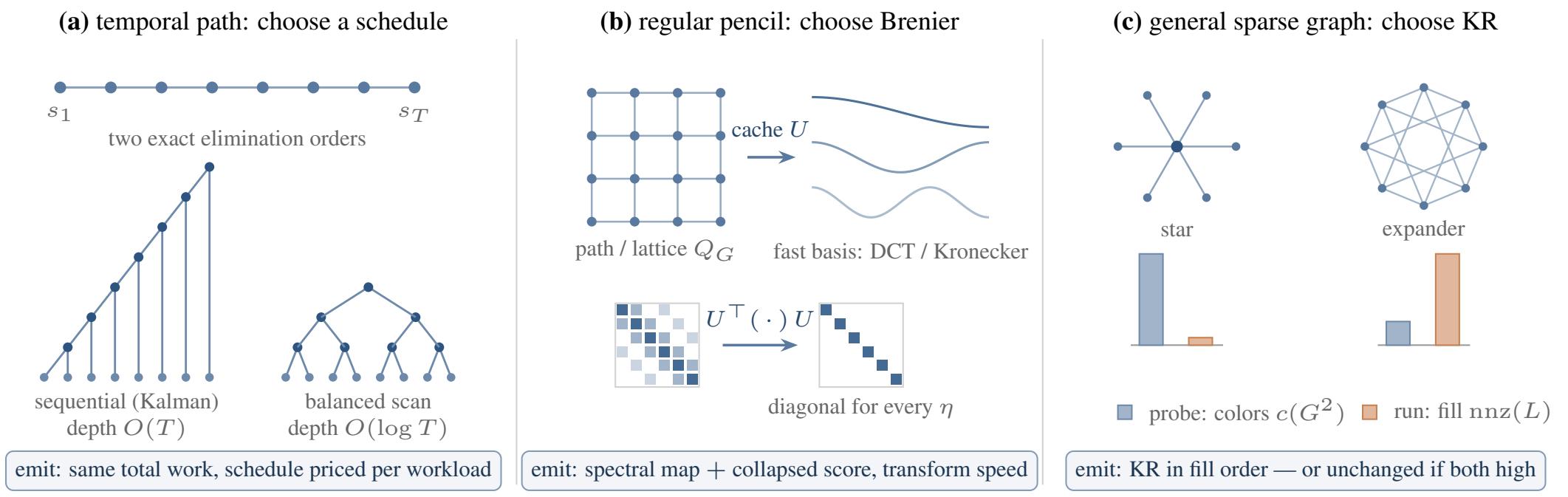

This appendix gives the protocol and complete results behind
Section~\ref{sec:gmrf}, and extends the chain and pentadiagonal cases to
distributions on general graph-indexed domains. Let $G=(V,E)$ be an undirected
graph with adjacency $A_G$, degree matrix $D_G$, and Laplacian $Q_G=D_G-A_G$;
Eq.~\eqref{eq:gmrf-model} uses $\kappa=0.2$, $u\sim\N(-1,0.7^2)$, and
$v\sim\N(-1.5,0.7^2)$, and the graph itself is supplied only inside the
black-box target used to generate and score data.

\paragraph{Precision recovery and factorization.}
Equation~\eqref{eq:gmrf-precision} has structural pattern $I+A_G$. Because its
diagonal is present, precision columns $i$ and $j$ share a structural row if
and only if their graph distance is at most two. The conflict graph for
Curtis--Powell--Reid compressed probes is therefore the square graph $G^2$;
the fixed greedy coloring used here needs $c(G^2)$ probe colors. This controls
colored precision assembly, but not sparse factorization. For a cached
fill-reducing order $P$, the backend computes
$P\Lambda P^\top=LDL^\top$ and applies the exact fiber map
\begin{equation}
 x_P=P(x-m),\qquad z_P=D^{1/2}L^\top x_P.
 \label{eq:sparse-layer}
\end{equation}
The $x$-to-$z$ Jacobian follows the filled elimination graph represented by
$L$, not generally the original graph $G$; cross-derivatives with respect to
the two core coordinates can be denser. Triangular maps cost
$O(\operatorname{nnz}(L))$. The collapsed score additionally needs selected
entries of $\Lambda^{-1}$ in
$\operatorname{tr}(\Lambda^{-1}\partial_\eta\Lambda)$; a Takahashi recursion
computes them on the filled graph without materializing the dense inverse.

\paragraph{Protocol.}
We test binary trees, square lattices, Delaunay graphs of random planar points,
3-regular expanders constructed as a cycle plus a random perfect matching, and
stars at $n\in\{64,256,576,1024\}$. Planar and expander cells use three graph
instances; deterministic families use one. Pattern detection uses the same
three random probes and held-out check as the rest of the paper. After one
warm call, timings are medians of seven complete calls. Symbolic minimum-degree
ordering is cached; every call performs colored oracle assembly, sparse numeric
factorization and solves, selected inversion, and the exact first-order
collapsed value and score. Dense Cholesky provides an independent numerical
check through $n=576$. Runs use the Linux arm64 environment of
Section~\ref{sec:dbn}.

\begin{figure}[t]
 \centering
 \includegraphics[width=\linewidth]{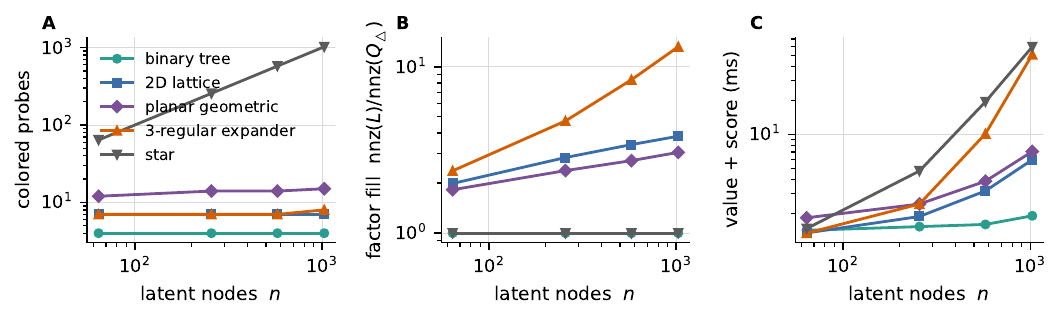}
 \caption{General-graph GMRF results. (A) Probe colors for $G^2$.
 (B) Factor fill relative to $Q_\triangle$, the lower triangle of the
 discovered precision. (C) Collapsed value-and-score time, including colored
 assembly, sparse factorization, and selected inversion. Random-family points
 are medians over three graph instances.}
 \label{fig:gmrf-suite}
\end{figure}

\begin{table}[t]
 \centering
 \footnotesize
 \setlength{\tabcolsep}{4pt}
 \caption{Sparse backend at $n=1024$. ``Factor'' is numeric LDL time after the
 cached symbolic order; ``score'' is the complete exact value-and-score call;
 ``memory'' is dense precision storage divided by sparse factor storage.
 Random-family entries are medians over three graph instances.}
 \label{tab:gmrf-suite}
 \fittable{%
 \begin{tabular}{@{}lrrrrr@{}}
  \toprule
  graph & colors & fill & factor (ms) & score (ms) & memory $\times$\\
  \midrule
  binary tree     & 4    & 1.00  & 0.21 & 1.90  & 228\\
  2D lattice      & 7    & 3.83  & 0.78 & 5.96  & 56\\
  planar geometric& 15   & 3.06  & 0.95 & 7.06  & 52\\
  3-regular expander & 8 & 13.19 & 2.22 & 51.10 & 20\\
  star            & 1024 & 1.00  & 0.21 & 59.87 & 228\\
  \bottomrule
 \end{tabular}}
\end{table}

Every one of the 36 graph/size instances recovers the precision pattern with
edge precision and recall exactly one. Dense and sparse log determinants agree
within $4.55\times10^{-13}$, transport round trips within
$2.22\times10^{-15}$, and held-out conditional-identity residuals within
$2.33\times10^{-10}$. At $n=1024$, one-time pattern discovery takes
0.13--0.30 seconds and cached symbolic ordering 3.3--6.1 milliseconds.
The timings in Table~\ref{tab:gmrf-suite} reflect two distinct costs. The
star has no fill but requires $n$ probe colors; the expander has few colors
but substantial fill. The lattice and planar graph are less costly on both
measures. Neither degree nor fill alone is sufficient to predict the cost of
the complete score evaluation.

\subsubsection{Shared-spectrum Brenier compilation}
\label{app:spectral}

The spectral branch uses an exact contract rather than attempting to infer a
matrix pencil from finitely many probes. Write
$Q_G+\kappa I=U\operatorname{diag}(\lambda)U^\top$,
$a=e^{-2u}$, $b=e^{-2v}$ and $d_i=a\lambda_i+b$. With
$\widehat y=U^\top y$, the emitted kernel evaluates
\begin{equation}
 \lcoll(u,v)=-nu-nv-\tfrac12b\|y\|^2
 +\tfrac12\sum_i\frac{b^2\widehat y_i^2}{d_i}
 +\tfrac n2\log(2\pi)-\tfrac12\sum_i\log d_i+\log p(u,v),
 \label{eq:spectral-collapsed}
\end{equation}
and its analytic two-dimensional score in $O(n)$ work. Its conditional mean is
$m=U\operatorname{diag}(b/d_i)\widehat y$ and its Brenier map is
$z=U\operatorname{diag}(\sqrt{d_i})U^\top(x-m)$. Orthonormal DCT-II evaluates
$U,U^\top$ on a path; applying it along both axes gives the Kronecker transform
on a square lattice. Both maps cost $O(n\log n)$ without storing $U$.

\begin{figure}[t]
 \centering
 \includegraphics[width=\linewidth]{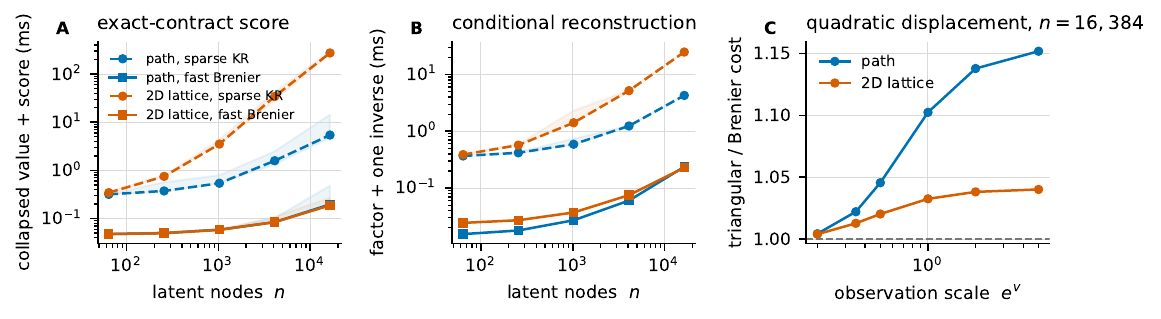}
 \caption{Exact-contract sparse KR versus fast-transform Brenier compilation.
 (A) Collapsed value-and-score time. (B) One conditional reconstruction,
 including the core-dependent factor or spectral mean. (C) Exact centered
 quadratic-cost ratio as observation noise weakens; Brenier is the unit line.}
 \label{fig:spectral-branch}
\end{figure}

The sparse comparator also receives the exact $Q_G$. It caches
a minimum-degree symbolic order, then performs sparse numeric $LDL^\top$ and
Takahashi selected inversion for every score; reconstruction performs the
factorization and one triangular inverse. This excludes the oracle assembly
cost in Figure~\ref{fig:gmrf-suite}, which favors the sparse comparator. A
cached dense
eigendecomposition is a third control through $n=1{,}024$. We use the same
$\kappa=0.2$, simulated scales $(e^u,e^v)=(0.35,0.20)$ and one fixed seed on
paths and square lattices for
$n\in\{64,256,1{,}024,4{,}096,16{,}384\}$. Timings are medians of 15 batches
after process-level FFT and BLAS initialization, on the CPU environment of
Section~\ref{sec:dbn}; bands in Figure~\ref{fig:spectral-branch} are batch
minima and maxima.

\begin{table}[t]
 \centering
 \footnotesize
 \setlength{\tabcolsep}{3pt}
 \caption{Shared-spectrum compilation. ``Score $\times$'' and ``recon.
 $\times$'' are sparse-KR time divided by fast-Brenier time. Dense build stores
 the full graph Fourier basis; dashes are builds intentionally not attempted.}
 \label{tab:spectral-branch}
 \fittable{\begin{tabular}{@{}llrrrrrr@{}}
\toprule
graph & $n$ & fast build & dense build & sparse score & fast score & score $\times$ & recon. $\times$\\
\midrule
path & 1,024 & 0.06 ms & 145.1 ms & 0.54 ms & 0.059 ms & 9.2 & 21.74\\
path & 16,384 & 0.23 ms & -- & 5.42 ms & 0.197 ms & 27.5 & 18.03\\
2D lattice & 1,024 & 0.06 ms & 157.5 ms & 3.52 ms & 0.059 ms & 60.1 & 37.88\\
2D lattice & 16,384 & 0.16 ms & -- & 274.59 ms & 0.188 ms & 1461.1 & 108.16\\
\bottomrule
\end{tabular}
}
\end{table}

Table~\ref{tab:spectral-branch} separates construction, score, and
reconstruction costs. At $n=1{,}024$, a cached dense spectrum takes
145--158\,ms to build and repays this through its $0.059$\,ms scores after
about 299 path or 45 lattice calls. Applying its basis remains slower than
the fast transforms: cached map pairs take 0.99--1.05\,ms, compared with
0.031--0.045\,ms. A shared spectrum
therefore helps collapsed inference even when a fast implementation of the
Brenier map is unavailable.

Across all ten cells, fast and sparse collapsed values agree relatively within
$8.4\times10^{-16}$ and scores within $2.4\times10^{-12}$; DCT diagonalization
and transport round trips agree within $4.4\times10^{-16}$ and
$3.6\times10^{-15}$. The transport-cost comparison is analytic, not sampled:
after removing the common mean shift, Brenier costs
$\operatorname{tr}(\Lambda^{-1})+n-2\operatorname{tr}(\Lambda^{-1/2})$,
whereas the triangular $LDL^\top$ coupling replaces the final trace by
$\sum_iD_{ii}^{-1/2}$. At the simulated noise scale the improvement is small but
strict; panel C shows that it grows as the graph prior controls more of the
fiber.

\paragraph{Matched spectral end-to-end sampling.}
We next hold the collapsed inference problem fixed and change only its exact
backend. On each graph at $n=1{,}024$, sparse KR and fast Brenier receive the
same Fisher-diagonal NUTS configuration, sampler seeds and initial positions:
four chains, 400 warmup iterations, 500 retained draws and maximum tree depth
six. Wall time charges backend construction, warmup and sampling, then backend
reconstruction of every one of the 2,000 retained field draws. A mode-centered
80-point Gauss--Legendre calculation in the exact spectrum supplies the
deterministic core reference. All 12 runs have zero sampling divergences and
worst $\rhat=1.020$; their worst mean error is $0.12$ reference standard
deviations and worst absolute log-SD error is $0.094$. Warmup divergences, not
tabled, total 50--53 per family and arm.

\begin{table}[t]
 \centering
 \footnotesize
 \setlength{\tabcolsep}{2.2pt}
 \caption{Matched exact-contract NUTS at $n=1{,}024$ (three-seed medians). Wall
 time includes construction, warmup, sampling, and 2,000 reconstructions;
 diagnostics show worst seeds and summed divergences.}
 \label{tab:spectral-nuts}
 \fittable{\begin{tabular}{llrrrrrrrr}
\toprule
Graph & map & wall (s) & min ESS & ESS/s & s/ESS & worst $\widehat R$ & div. & mean $z$ & pred./obs. \\
\midrule
path (1,024) & sparse KR & 20.05 & 523 & 26.35 & 0.038 & 1.019 & 0 & 0.03 & 0.94 \\
 & fast Brenier & 2.55 & 534 & 203.31 & 0.005 & 1.008 & 0 & 0.08 & 0.47 \\
\addlinespace
2D lattice (1,024) & sparse KR & 140.45 & 581 & 4.14 & 0.242 & 1.020 & 0 & 0.08 & 0.97 \\
 & fast Brenier & 2.94 & 495 & 157.69 & 0.006 & 1.016 & 0 & 0.12 & 0.46 \\
\bottomrule
\end{tabular}
}
\end{table}

The two backends do similar statistical work: sparse/fast median
active-gradient counts are 18,362/18,169 on the path and 20,920/20,124 on the
lattice, and fast/sparse median minimum ESS per active gradient is $0.99$ and
$0.87$. Fast Brenier nevertheless reduces median seconds per minimum ESS by
$7.7\times$ on the path and $38.1\times$ on the lattice
(Table~\ref{tab:spectral-nuts}). The difference is in latency: calibrated
chain-batch scores are $15.0\times$ and $94.4\times$ faster, and conditional
draws $62.0\times$ and $103\times$ faster. The workload estimate accounts for
94--97\% of sparse runtime and 46--47\% of fast runtime, where sampler and
language overhead become appreciable. We stopped the sampling sweep at
$n=1{,}024$ under the extension rule: the first sparse lattice run already
took 141 seconds. Only the kernel timings extend to $n=16{,}384$.

\paragraph{Oracle-emitted end-to-end sampling.}
The three-arm comparison of Section~\ref{sec:gmrf} fixes $n=256$ and graph
instance 0 per family, with
three sampler seeds, four chains, 500 retained draws and maximum tree depth six.
The two fixed arms use 400 warmup iterations and the Fisher-diagonal NUTS
implementation used by the real-model checks in Appendix~\ref{app:census}; the
learned arm uses nutpie's flow
adaptation with 800. Because that adapter needs a differentiable log density in
its own framework, the raw target is transcribed independently for it and checked
against the target driving the other two arms at random probe points before any
sampling; the worst relative disagreement is $4.6\times10^{-16}$. All three arms
therefore see the same simulated data, graph instance and quadrature reference.
Flow rows count sampled-target gradients only, so they omit the cost of fitting
and evaluating the map itself, which wall time does charge. Divergences count the
sampling phase, as elsewhere in the paper; all three arms accumulate warmup
divergences in the tens, fewest for the flow arm.

Table~\ref{tab:gmrf-nuts} gives the values behind
Figure~\ref{fig:gmrf-nuts}. On the planar graph, collapse and learning achieve
similar ESS/s: the collapsed score requires 15 colors, whereas the flow uses
fewer gradients but incurs fitting costs. On the star, every collapsed
gradient requires 256 colors and one seed still reaches only
$\rhat=1.060$. The coloring cost is available before sampling and suffices
to reject this oracle-based implementation.

\begin{table}[ht]
 \centering
 \footnotesize
 \setlength{\tabcolsep}{4.5pt}
 \caption{Three-arm sampling at $n=256$, medians over three sampler seeds
 (worst $\rhat$ and total divergences are over seeds). ``Gate'' is
 core rank-$\rhat\le1.05$ in every seed. Rows are target-equivalent and exclude
 flow evaluation; time is end to end, including discovery and reconstruction.}
 \label{tab:gmrf-nuts}
 \fittable{\begin{tabular}{@{}llcrrrrr@{}}
\toprule
graph & arm & gate & worst $\rhat$ & div.\ & min core ESS & rows/ESS & time (s)\\
\midrule
binary tree & raw NUTS & \checkmark & 1.029 & 0 & 121 & 546 & 12 \\
 & learned flow & \checkmark & 1.006 & 0 & 1194 & 65 & 133 \\
 & sparse RB-NUTS & \checkmark & 1.005 & 0 & 686 & 371 & 26 \\
\addlinespace
2D lattice & raw NUTS & \emph{fail} & 1.136 & 0 & 35 & 1,959 & 13 \\
 & learned flow & \checkmark & 1.011 & 0 & 549 & 147 & 144 \\
 & sparse RB-NUTS & \checkmark & 1.006 & 0 & 547 & 960 & 52 \\
\addlinespace
planar geometric & raw NUTS & \checkmark & 1.045 & 0 & 149 & 468 & 14 \\
 & learned flow & \checkmark & 1.002 & 0 & 1096 & 72 & 161 \\
 & sparse RB-NUTS & \checkmark & 1.013 & 0 & 481 & 1,944 & 66 \\
\addlinespace
3-regular expander & raw NUTS & \emph{fail} & 1.108 & 0 & 122 & 557 & 12 \\
 & learned flow & \checkmark & 1.009 & 0 & 861 & 95 & 149 \\
 & sparse RB-NUTS & \checkmark & 1.017 & 0 & 477 & 1,067 & 60 \\
\addlinespace
star & raw NUTS & \emph{fail} & 2.169 & 0 & 9 & 7,415 & 13 \\
 & learned flow & \emph{fail} & 1.081 & 10 & 116 & 653 & 146 \\
 & sparse RB-NUTS & \emph{fail} & 1.060 & 9 & 178 & 152,151 & 300 \\
\bottomrule
\end{tabular}
}
\end{table}

The ridge $\kappa I$ in Eq.~\eqref{eq:gmrf-model} avoids the intrinsic CAR null
mode. A literal ICAR can instead impose a sum-to-zero chart or acquire a
positive-definite posterior conditional through anchored Gaussian
observations; the layer requires the resulting $\Lambda(\eta)$ to be positive
definite.

\FloatBarrier
\subsection{Temporal path schedules and sampler scaling}
\label{app:scaling}

\begin{figure}[ht]
 \centering
 \includegraphics[width=\linewidth]{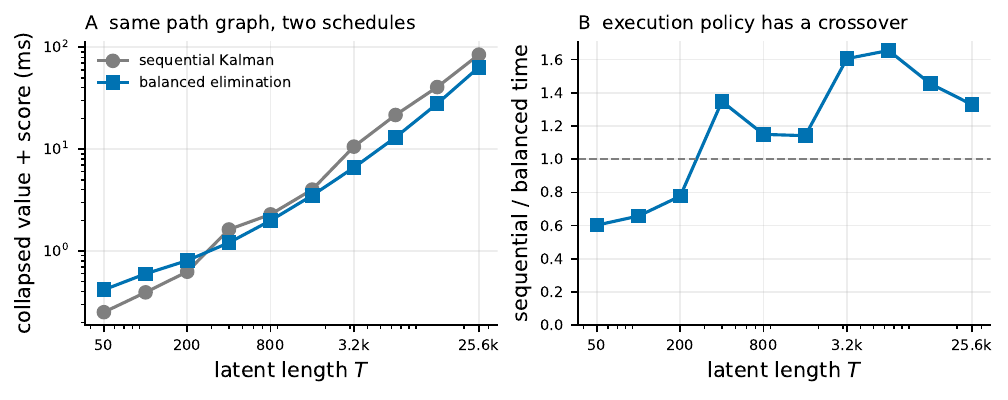}
 \caption{Compiling one recovered temporal path into two exact CPU schedules.
 (A) Median of 15 post-compilation calls to the collapsed value and its
 eight-dimensional score. (B) Speedup of balanced elimination over the
 sequential Kalman scan. The sweep isolates the recurrent emitted kernel; it
 does not rerun the current quadratic-memory discovery front end at each $T$.}
 \label{fig:dbn-associative}
\end{figure}

For the schedule comparison we simulate one nested dataset at lengths doubling
from $T=50$ through $T=25{,}600$, with fixed truth, sensor spacing and seed.
Timings are medians of 15 calls after separate JIT compilation, in float64 JAX
on ten arm64 CPU cores with no GPU. Correctness is checked at the truth and four
fixed perturbations; the worst relative discrepancy is $6.8\times10^{-15}$ in
value and $2.9\times10^{-12}$ in score. Figure~\ref{fig:dbn-associative} shows
the operation-dependent crossover.
Balanced elimination loses for score calls through $T=200$, wins from $T=400$,
and remains $1.33\times$ faster at $T=25{,}600$ (51,200 latent coordinates).
For values alone, sequential execution wins throughout. Score compilation costs
$1.2$--$4.0$ seconds for the balanced program versus $0.20$--$0.33$ seconds for
the sequential one, yielding the reported workload-dependent break-even points.

\begin{figure}[ht]
 \centering
 \includegraphics[width=\linewidth]{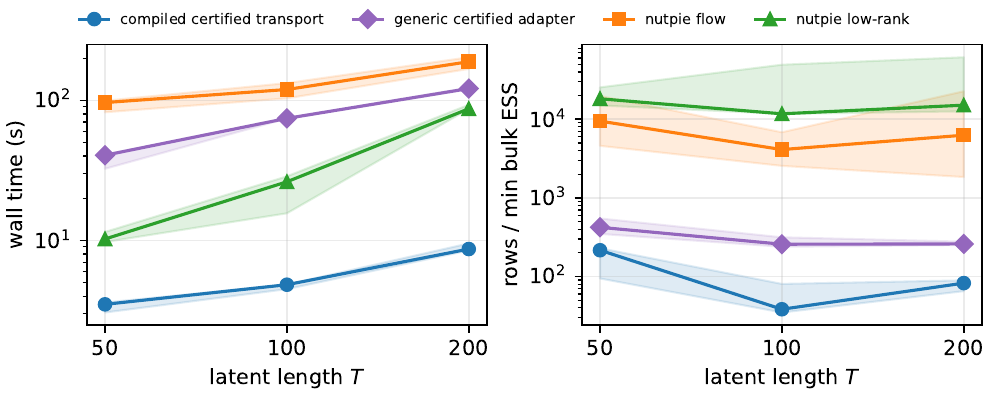}
 \caption{Four-way DBN scaling. Lines are three-seed medians and bands are
 min--max. Both axes are logarithmic. The compiled transport scales from
 3.50 to 8.69 seconds over $T=50$--$200$. Rows per minimum ESS for
 nonconverged learned runs are failure diagnostics, not valid efficiency
 estimates.}
 \label{fig:dbn-scaling}
\end{figure}

Figure~\ref{fig:dbn-scaling} plots the four arms of Section~\ref{sec:dbn}
against latent length and Table~\ref{tab:diagnostics} gives their per-cell
diagnostics; the curve is placed here because the timing comparison is the least
surprising of the three DBN results. The generic adapter has no divergences,
worst $\rhat\le1.036$, and median minimum ESS $162,385,380$ across the three
lengths; the compiled arm has no divergences at $T=100,200$ and worst
$\rhat\le1.048$.

Two sensitivity runs support the compiled arm beyond the main protocol.
Repeating the anomalous $T=50$, seed-2 run with 800 warmup and 600 draws reduces
divergences from 12 to 1 and $\rhat$ from 1.048 to 1.013. At $T=400$ (808
dimensions), three runs with that longer protocol have zero divergences, worst
$\rhat=1.023$, median minimum ESS 293, and median wall time 53.54 seconds. Both
are kept out of the timing curve because their budget differs.

\begin{table}[ht]
 \centering
 \footnotesize
 \setlength{\tabcolsep}{4.2pt}
 \caption{Worst rank-$\rhat$ and total divergences across three seeds. Each
 cell is $\max\rhat$/divergences. The learned methods often fail standard
 convergence criteria, especially at $T=200$.}
 \label{tab:diagnostics}
 \fittable{%
 \begin{tabular}{@{}rcccc@{}}
  \toprule
  $T$ & compiled & generic adapter & learned flow & low-rank\\
  \midrule
   50 & 1.048/12 & 1.036/0 & 1.441/5 & 1.748/13\\
  100 & 1.037/0  & 1.020/0 & 1.162/1 & 1.940/8\\
  200 & 1.040/0  & 1.033/0 & 1.963/2 & 2.157/143\\
  \bottomrule
 \end{tabular}}
\end{table}

\subsection{Separating compilation and metric adaptation}
\label{app:attribution}

The compiled transport is $10.8$--$15.0\times$ faster than the generic adapter
on the same conditional map. Two changes contribute to this difference.
Compilation replaces density evaluation and differentiation through the map
with a collapsed value-and-score calculation. The compiled model also
receives nutpie's diagonal mass-matrix adaptation, while the transform adapter
holds its metric fixed. Since Proposition~\ref{prop:factorization}
standardizes only the Gaussian block, adaptation can still help the core.

We compare three implementations of Eq.~\eqref{eq:compiled-score}.
A is the compiled model with diagonal adaptation; B uses the compiled score
through an identity transform adapter with a fixed metric; C is the generic
adapter with the same fixed metric. Comparing C with B isolates score
evaluation, and comparing B with A isolates adaptation. Each uses three seeds,
two chains, 100 warmup iterations, 300 retained draws, and maximum tree depth 7.

\begin{table}[ht]
 \centering
 \footnotesize
 \setlength{\tabcolsep}{4pt}
 \caption{Compilation and metric adaptation, medians over three seeds.
 C/B compares score implementations at a fixed metric; B/A compares fixed
 and adapted metrics with the compiled score.}
 \label{tab:attribution}
 \fittable{\begin{tabular}{@{}llrrrrrr@{}}
\toprule
$T$ & arm & steps/draw & grad.\ rows & $\mu$s/row & time (s) & min ESS & s / min ESS\\
\midrule
$50$ & A: compiled, diagonal metric & 14.7 & 12,268 & 187 & 2.22 & 70 & 0.040\\
 & B: compiled, identity metric & 76.0 & 62,701 & 254 & 15.80 & 171 & 0.089\\
 & C: generic adapter, identity metric & 67.5 & 56,570 & 423 & 23.84 & 218 & 0.108\\
\multicolumn{2}{@{}l}{\quad attribution} & \multicolumn{6}{l}{compilation $1.67\times$ per row (C/B); metric $5.1\times$ in rows (B/A); end to end $10.8\times$, or $2.7\times$ per ESS}\\
\addlinespace[2pt]
$100$ & A: compiled, diagonal metric & 14.3 & 11,587 & 320 & 3.76 & 246 & 0.017\\
 & B: compiled, identity metric & 127.0 & 98,943 & 352 & 34.89 & 348 & 0.101\\
 & C: generic adapter, identity metric & 127.0 & 99,161 & 573 & 56.46 & 345 & 0.166\\
\multicolumn{2}{@{}l}{\quad attribution} & \multicolumn{6}{l}{compilation $1.63\times$ per row (C/B); metric $8.5\times$ in rows (B/A); end to end $15.0\times$, or $9.8\times$ per ESS}\\
\addlinespace[2pt]
$200$ & A: compiled, diagonal metric & 15.0 & 12,550 & 588 & 7.51 & 167 & 0.043\\
 & B: compiled, identity metric & 127.0 & 98,780 & 632 & 62.46 & 372 & 0.169\\
 & C: generic adapter, identity metric & 127.0 & 98,986 & 914 & 90.90 & 345 & 0.263\\
\multicolumn{2}{@{}l}{\quad attribution} & \multicolumn{6}{l}{compilation $1.44\times$ per row (C/B); metric $7.9\times$ in rows (B/A); end to end $12.1\times$, or $6.1\times$ per ESS}\\
\bottomrule
\end{tabular}
}
\end{table}

With the metric fixed, compilation reduces time per gradient by
$1.44$--$1.67\times$ (Table~\ref{tab:attribution}). Metric adaptation has a
larger effect on gradient counts: B and C reach the 127-step tree-depth limit
at $T\ge100$, compared with 14--15 steps for A, and use $5.1$--$8.5\times$
more gradients per retained draw. Their longer trajectories also produce
higher minimum ESS. Accounting for this narrows the overall advantage of A
from $10.8$--$15.0\times$ in runtime to $2.7$--$9.8\times$ in seconds per
minimum ESS. The spread of the latter range across lengths reflects the noise
in a minimum-ESS estimate from 600 draws and three seeds. The runtime
difference should therefore be attributed to both compilation and adaptation
of the remaining core. Compiling to an ordinary density keeps the standard
adaptation path available for that core.

\FloatBarrier
\subsection{Learning only the residual core: the
\texorpdfstring{$(d_g,d_h)$}{(dg, dh)} sweep}
\label{app:core}

Corollary~\ref{cor:objectives} says a learned flow should be fitted after the
certificate rather than asked to rediscover the Gaussian state geometry. This
appendix gives the two experiments behind the result summarized in
Section~\ref{sec:experiments}: a matched
capacity--budget comparison at one split, and a sweep of the split itself.

\subsubsection{One split, matched capacity and budget}

At $T=100$ (208 coordinates) the full arm learns all 208 coordinates while the
hybrid arm composes the compiled certificate with the same flow acting only on
the eight-dimensional core, over matched capacities and warmup budgets. The
experiment uses one masked flow family (width 16, depth 1) at two or eight
layers and warmup budgets in $\{100,300,800\}$, with seeds $0,1,2$, two chains and
300 retained draws per cell. Relative Fisher divergence is evaluated on 512
held-out draws from four independent compiled chains (zero divergences, worst
$\rhat=1.010$, minimum bulk ESS 854), so the held-out objective is scored against
draws no learned arm contributed to. Before any learning, the certificate alone
reduces that objective from $9.7\times10^5$ in raw coordinates to
$1.83\times10^3$ on the factorized target, so the learned comparison measures how
efficiently each flow fits the remaining mismatch.

\begin{figure}[ht]
 \centering
 \includegraphics[width=\linewidth]{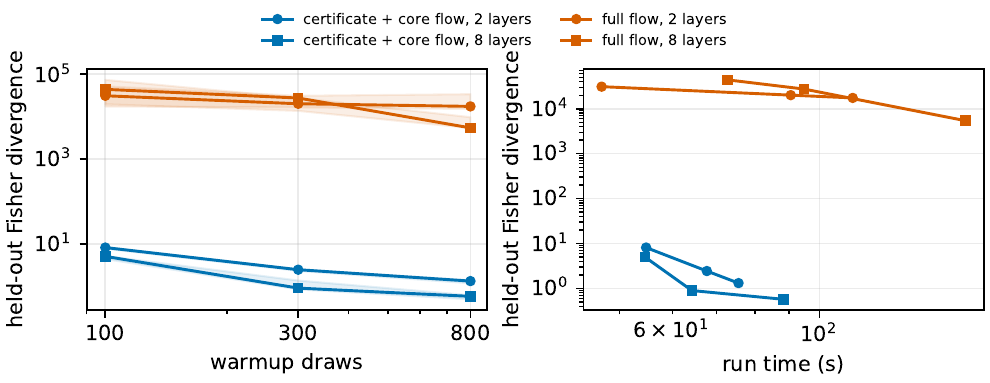}
 \caption{Matched residual-flow experiment at $T=100$: the same two- or
 eight-layer flow trained on all 208 raw coordinates (full flow) or on the
 eight-dimensional core after certification. Held-out Fisher divergence against
 warmup budget and run time; three-seed medians, logarithmic axes.}
 \label{fig:hybrid-flow}
\end{figure}

Across the six capacity--budget pairs in Figure~\ref{fig:hybrid-flow}, the
hybrid median Fisher divergence is $3.8\times10^3$--$3.0\times10^4$ lower
than that of the full flow. Learning on the core reduces the parameter count
by $14.1\times$. At the largest setting, the divergences are 0.574 and
$5.40\times10^3$, reached in 88.2 and 166.1 seconds, respectively. All 18
hybrid runs have zero divergences, worst $\rhat=1.034$, and minimum ESS at
least 170. The full-flow runs reach $\rhat=2.32$ and minimum ESS as low as
2.5. We therefore use the independent held-out Fisher objective for the
learning comparison; ESS from the unconverged runs is only diagnostic.

\subsubsection{Sweeping the split}

Section~\ref{sec:experiments} reports the largest matched setting at the split
$(d_g,d_h)=(200,8)$; the experiment above also varies capacity and warmup.
Because the pushforward factorizes, the hybrid arm's
objective \emph{is} the core objective: collapse cannot degrade the core's
geometry, and there is no approximation term to trade off. The prescription can
therefore fail only through the core, if what is left is not cheaply learnable
per unit cost---and every core the paper starts from is otherwise benign: the
DBN's eight coordinates are linear-Gaussian hyperparameters, the graph field's is
two-dimensional. The untested axes are core dimension and core non-Gaussianity.
This appendix sweeps them on the DBN;
Appendix~\ref{app:graphcore} repeats the sweep on the graph fibers, where the
exponent turns out to depend on the fiber.

\paragraph{Inflating the core at a fixed block.}
We refine the DBN's $y$-sensor log-scale into $K$ contiguous time segments, so
$\eta$ gains $K-1$ coordinates and $d_h=7+K$ while the state block stays at
$d_g=2T$. Segments enter the banded layer and the Kalman scan only through
per-time sufficient statistics, so the block algebra costs the same at every core
dimension, and $K=1$ is the model of Section~\ref{sec:experiments}
itself---bit-identical collapsed density, verified against that implementation.
Refining a \emph{fixed} dataset rather than adding sensors holds the target's
information fixed: at $T=100$, each of $K=1,9,25$ scales is informed by $100$,
$11$ and $4$ observations, so the core grows in dimension and in
non-Gaussianity together, which is the unfavorable direction. The data, states, and generating process remain fixed, so the comparison
isolates $d_h$. Held-out identity residuals for the emitted contract stay below
$1.2\times10^{-15}$ relative at every $K$.

\paragraph{Protocol.}
Both arms use the flow family, chains, draws and tree depth of the matched
experiment above, at the setting most favorable to the baseline---eight layers,
800 warmup---with five seeds per cell. Because ``matched capacity'' is not a
fixed parameter ratio once the learned dimension changes, we report the realized
trainable-parameter count per cell rather than quoting one ratio. The held-out
objective is scored on 512 draws from an independent compiled chain per target,
so no learned arm contributes to its own evaluation.

\paragraph{Reading the separation.}
The ratio of the two arms' trained objectives factorizes exactly. Write $\F$ for
a held-out value of $\FD$, so that $\F_{\mathrm{raw}}$ and $\F_{\mathrm{cert}}$
are the two coordinate systems scored under the identity map, and let
$G=\F_{\mathrm{identity}}/\F_{\mathrm{trained}}$ be how much an arm's training
improves on its own starting point. Then
\begin{equation}
 \underbrace{\frac{\F_{\mathrm{full}}}{\F_{\mathrm{hybrid}}}}_{\text{separation}}
 =\underbrace{\frac{\F_{\mathrm{raw}}}{\F_{\mathrm{cert}}}}_{\text{free, before training}}
 \times\underbrace{\frac{G_{\mathrm{hybrid}}}{G_{\mathrm{full}}}}_{\text{relative training gain}}.
 \label{eq:separation-decomposition}
\end{equation}
The first factor is the reduction supplied by the certificate at zero
training cost. The second measures whether the smaller learning problem is
easier to solve. A benefit due solely to dimension reduction would change
only the first factor as $d_h$ varies.

\paragraph{Results.}
At $d_h=8$ the hybrid's median held-out divergence is $0.52$ against
$5.45\times10^4$ for the full flow, a separation of $1.06\times10^5$; at
$d_h=16$ it is $1.54$ against $1.54\times10^4$, a separation of
$9.99\times10^3$; at $d_h=32$ it is $3.97$ against $6.75\times10^3$, a separation
of $1.70\times10^3$. Quadrupling the core costs a factor of $62$, and over these
three points the separation scales as $d_h^{-2.98}$. Seed ranges do not overlap
between adjacent core sizes---$[3.8,13.7]\times10^4$, $[0.4,1.6]\times10^4$ and
$[970,2200]$---so the decay is larger than the seed-to-seed spread. Realized
parameter counts make ``matched capacity'' explicit: the core flow holds
$19{,}776$, $30{,}160$ and $50{,}928$ trainable parameters against the full
flow's $279{,}376$, $289{,}760$ and $310{,}528$, so the reduction falls from
$14.1\times$ to $6.1\times$. Thus, matching the architecture does not hold the parameter ratio fixed
across the sweep.

Equation~\eqref{eq:separation-decomposition} locates the decay. The free head
start falls only as $d_h^{-0.77}$, from $2{,}313$ to $793$; these ratios inherit
a $30$--$60\%$ Monte Carlo error from the heavy-tailed per-draw cost of the raw
coordinate system, so they are good to roughly a factor of two. The relative
training gain falls as $d_h^{-2.21}$, from $45.7$ to $2.14$, because the hybrid's
own gain collapses from $3{,}488$ to $426$ while the full flow's \emph{improves}
from $76$ to $199$. Three quarters of the decay exponent is therefore the term
that measures whether the residual is easy to fit, not the term that measures
what collapse removes---which is what Corollary~\ref{cor:objectives} predicts,
since collapse is exact at every $d_h$ and only the premise that what remains is
cheaply learnable can weaken. Extrapolating the fitted exponent to parity puts
the crossover near $d_h\approx390$ at $d_g=200$: on this family the core would
have to be about twice the block before compiling stopped paying. The
extrapolation rests on three points, and Appendix~\ref{app:graphcore} shows that
the exponent itself is a property of the fiber---between $-0.31$ and $-0.59$ on
the graph families---so the crossover is specific to this state-space target and
should not be read as a general threshold.

\paragraph{Reliability does not follow the objective.}
All 15 hybrid runs have zero divergences, worst $\rhat=1.047$, and minimum bulk
ESS at least 186, at every core size. The 15 full-flow runs accumulate 2, 15 and
116 divergences as $d_h$ grows, reach $\rhat=1.42$, and never exceed minimum ESS
$9.3$. Wall time is flat too, $70$--$75$ seconds against $121$--$134$. As the core grows, the advantage in the training objective decreases,
while divergence-free sampling, usable ESS, and the $1.7\times$ wall-time
advantage persist throughout the tested range. The full-flow arm degrades.

\paragraph{The complementary axis.}
Inflating the block at a fixed core runs $T$ from 5 to 200 at $K=1$, taking $d_g$
from 10 to 400 against a fixed $d_h=8$. This is the friendly direction, and it
ends near a regime the paper flags as a refusal: at $T=5$ the whole target is
18-dimensional and a flow can learn all of it cheaply. Over three seeds per cell
the separation on this axis is not monotone in $d_g$---$1.6\times10^5$,
$3.9\times10^5$, $224$, $312$ and $8.3\times10^3$ at $d_g=10,20,50,100,400$---
because it is dominated by the free-reduction factor, that is, by how badly the
raw coordinate system happens to behave at each $T$, and that factor carries
$26$--$99\%$ Monte Carlo error here. We therefore read no trend off this axis and
rely on the core axis, where the factors separate cleanly. The centered
eight-schools posterior anchors the small-block end from real data, with
$(d_g,d_h)=(9,1)$: its $(\theta,\mu)$ block is Gaussian given $\log\tau$ even
though the $\tau$ prior is heavy-tailed, because the prior touches only the core,
and there the learned flow wins.

\FloatBarrier
\subsection{Inflating the core of a graph-indexed field}
\label{app:graphcore}

Appendix~\ref{app:core} inflates the state-space core. This appendix gives the
corresponding protocol, front-end checks, and complete results on
the graph-indexed field of Section~\ref{sec:gmrf}.

\paragraph{Model and design.}
We assign a separate observation log-scale to each of $K$ graph regions:
$\Lambda(\eta)=e^{-2u}(Q_G+\kappa I)+\mathrm{diag}(e^{-2v_{r(i)}})$,
where $\eta=(u,v_1,\dots,v_K)$ and $d_h=K+1$. This changes the diagonal
entries but preserves the precision pattern $I+A_G$. Regions are contiguous
segments of a fixed breadth-first ordering. For each family, we simulate one
dataset at a common observation scale and refine the partition while holding
the data fixed. At $n=256$ and $K=1,4,16,64$, each scale has 256, 64, 16,
or four observations. The $K=1$ model is Eq.~\eqref{eq:gmrf-model}.

We use binary trees, square lattices, 3-regular expanders, and stars. Both
learning methods use the eight-layer, 800-warmup configuration of
Appendix~\ref{app:core}, the setting most favorable to the full-flow
baseline, with five sampler seeds (four in seven of the 32
arms where the grid stopped early). An independent compiled chain supplies
512 reference draws per target.

\paragraph{Structural checks.}
Discovery on an independent PyTorch transcription recovers $I+A_G$ for all
16 targets. Colors, bandwidth, and fill are unchanged across $K$; the
color/fill pairs are $4/1.00$ (tree), $7/2.85$ (lattice), $8/4.63$
(expander), and $256/1.00$ (star). Held-out identity residuals are below
$1.4\times10^{-11}$ nats. PyTorch and JAX values agree to
$7\times10^{-16}$ relative, and the $K=1$ density differences agree with
the original model to $1.3\times10^{-14}$. Inflating the core therefore
changes the learning problem and nothing else.

\begin{table}[ht]
 \centering
 \footnotesize
 \setlength{\tabcolsep}{4pt}
 \caption{Core inflation on four graph-indexed fibers at $d_g=256$, five seeds
 per cell and four in seven of the 32 arms. Colors and fill are the front-end invariants, identical at every
 $d_h$; ``free'' and ``gain'' are the two factors of
 Eq.~\eqref{eq:separation-decomposition}; ``s/ESS speedup'' is the hybrid's
 median advantage in seconds per minimum bulk ESS.}
 \label{tab:graphcore}
 \fittable{\begin{tabular}{@{}lrrrrrrrrrr@{}}
\toprule
family & $c(G^2)$ & fill & $d_h$ & obs per & \multicolumn{2}{c}{held-out $\FD$, median} & \multicolumn{3}{c}{separation $=$ free $\times$ gain} & s/ESS\\
\cmidrule(lr){6-7}\cmidrule(lr){8-10}
 & & & & scale & full & hybrid & total & free & gain & speedup\\
\midrule
binary tree & 4 & 1.00 & 2 & $256$ & $16.1$ & $0.0111$ & $1.45\times10^{3}$ & $33.1$ & $44$ & $1.70$\\
 & &  & 5 & $64$ & $43.4$ & $0.0281$ & $1.55\times10^{3}$ & $53.1$ & $29$ & $6.63$\\
 & &  & 17 & $16$ & $161$ & $0.158$ & $1.02\times10^{3}$ & $54.2$ & $19$ & $16.05$\\
 & &  & 65 & $4$ & $364$ & $0.674$ & $541$ & $35.6$ & $15$ & $3.52$\\
\multicolumn{4}{@{}l}{\quad fitted $d_h^{-0.30}$ over 4 core sizes} & \multicolumn{7}{r@{}}{}\\
\addlinespace[1.5pt]
2D lattice & 7 & 2.85 & 2 & $256$ & $72$ & $0.0612$ & $1.18\times10^{3}$ & $95$ & $12$ & $6.79$\\
 & &  & 5 & $64$ & $70.7$ & $0.179$ & $395$ & $69.8$ & $5.7$ & $6.99$\\
 & &  & 17 & $16$ & $121$ & $0.378$ & $319$ & $60.6$ & $5.3$ & $6.63$\\
 & &  & 65 & $4$ & $255$ & $1.43$ & $178$ & $50.5$ & $3.5$ & $4.89$\\
\multicolumn{4}{@{}l}{\quad fitted $d_h^{-0.49}$ over 4 core sizes} & \multicolumn{7}{r@{}}{}\\
\addlinespace[1.5pt]
3-regular expander & 8 & 4.63 & 2 & $256$ & $57.5$ & $0.0509$ & $1.13\times10^{3}$ & $41.6$ & $27$ & $3.56$\\
 & &  & 5 & $64$ & $54.9$ & $0.0986$ & $557$ & $53$ & $11$ & $4.21$\\
 & &  & 17 & $16$ & $117$ & $0.42$ & $279$ & $54.8$ & $5.1$ & $5.67$\\
 & &  & 65 & $4$ & $252$ & $1.8$ & $140$ & $47.2$ & $3$ & $4.93$\\
\multicolumn{4}{@{}l}{\quad fitted $d_h^{-0.59}$ over 4 core sizes} & \multicolumn{7}{r@{}}{}\\
\addlinespace[1.5pt]
star & 256 & 1.00 & 2 & $256$ & $51.3$ & $0.107$ & $481$ & $61.3$ & $7.9$ & $16.07$\\
 & &  & 5 & $64$ & $43.4$ & $0.239$ & $181$ & $60.7$ & $3$ & $9.92$\\
 & &  & 17 & $16$ & $172$ & $0.106$ & $1.63\times10^{3}$ & $59.3$ & $27$ & $10.67$\\
 & &  & 65 & $4$ & $364$ & $0.301$ & $1.21\times10^{3}$ & $39.8$ & $30$ & $3.24$\\
\multicolumn{4}{@{}l}{\quad refused in advance (256 colors); worst deviation from any power law $2.6\times$} & \multicolumn{7}{r@{}}{}\\
\bottomrule
\end{tabular}
}
\end{table}

\begin{figure}[t]
 \centering
 \includegraphics[width=\linewidth]{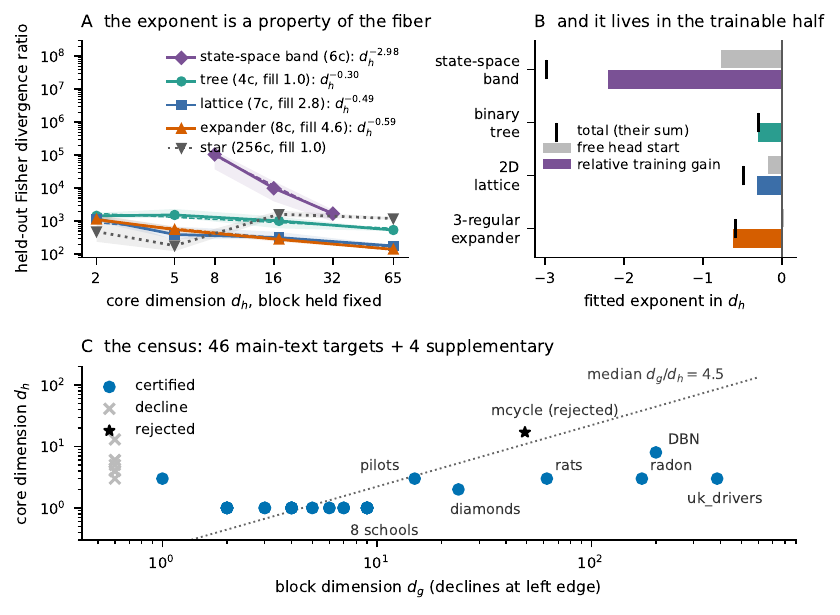}
 \caption{Complete residual-learning and census figure. Panels (A) and (B)
 summarize the state-space and graph core sweeps; panel (C) places the 46 main
 targets and four supplementary targets on the block/core plane.}
 \label{fig:decision-census}
\end{figure}

\paragraph{Decomposing the decay.}
Corollary~\ref{cor:objectives} at $S_\phi=\mathrm{id}$ makes the pre-training
objective equal to $\FD(p(\eta)\|\varphi_k)$, so the two factors of
Eq.~\eqref{eq:separation-decomposition} can be read separately. Measured over
the 16 cells, the raw objective is flat in the core dimension
($d_h^{-0.05}$ to $d_h^{+0.12}$ by family) and so is the certified one
($d_h^{+0.07}$ to $d_h^{+0.12}$), so the free head start is
$d_h^{0\pm0.17}$---consistent with additivity plus weak identification, since at
$K=64$ each region scale sees four observations and its posterior is
prior-dominated, hence nearly Gaussian, hence contributes almost nothing to the
core's own divergence. Because the fits are log-linear the two exponents add
exactly, and the split is visible in Figure~\ref{fig:decision-census}(B): the trainable
half carries $65$--$105\%$ of each fiber's decay ($-0.31$ of $-0.30$ on the tree,
$-0.32$ of $-0.49$ on the lattice, $-0.62$ of $-0.59$ on the expander, $-2.21$ of
$-2.98$ on the state-space band), the free head start supplying the small
remainder and being flat to within noise on two of the four. On the tree, for
instance, $G_{\mathrm{hybrid}}=\F_{\mathrm{cert}}/\F_{\mathrm{hybrid}}$ falls
from $3.2\times10^{4}$ at $d_h=2$ to $8.4\times10^{2}$ at $d_h=65$ while the
target it fits gets no harder intrinsically. The propositions locate this effect in the core. They exclude any
contribution from the block, and Proposition~\ref{prop:robustness} bounds the
contribution of certificate error at $2\eps\le3\times10^{-11}$ nats, ten orders
below the smallest measured objective. The remaining effect concerns
estimation in the $k$-dimensional core.

\paragraph{Graph structure and core dependence.}
If the fiber can only act through $p(\eta)$, the graph must reach the exponent
through the one term that carries it, $-\tfrac12\log|\Lambda(\eta)|$, whose
derivative $\operatorname{tr}(\Lambda^{-1}\partial_\eta\Lambda)$ runs over the
\emph{filled} elimination graph: more fill, more non-negligible entries of
$\Lambda^{-1}$, and a less separable core at the same $d_h$. We assess this dependence without sampling by finding the mode of
$\lcoll$ and evaluating the off-diagonal entries of the core information
$-\nabla^2\lcoll$ there. At $d_h=17$ the mean absolute off-diagonal of the
correlation-normalized information is $0.023$ (tree), $0.029$ (lattice) and
$0.034$ (expander), ordered as the fill is and as the exponents are, and the
fill-free tree has both the least coupled and the best conditioned core at every
$K$ (condition number $3.1$--$4.5$ against $5.9$--$11.4$). But the support is
partial and we report it as such: at $d_h=5$ and $d_h=65$ the lattice and the
expander are within noise of each other, their conditioning orders the two the
\emph{other} way, three families cannot separate fill from degree or spectral
gap, and the star---most coupled of all at $d_h=5$ and worst conditioned
everywhere---has no decay at all. Coupling therefore looks like part of the
mechanism rather than the whole of it, and the fill-ordered exponents remain
three points, not a law.

\paragraph{Results.}
Table~\ref{tab:graphcore} gives every cell. The separation decays with $d_h$ on
the three families the pass fires on, at exponents that are far from the DBN's
$-2.98$ and ordered by elimination fill, and the decomposition locates the decay
in the same place as on the DBN: the free head start is roughly flat in $d_h$
while the relative training gain falls. The star is the control. The
oracle-backend policy refuses it in advance at 256 colors, and its measured separation is non-monotone
in $d_h$, so a practitioner who read only the objective would collapse a fiber
whose oracle-driven execution costs $n$ probe passes per gradient. Two caveats
belong with these numbers. The collapsed arm here evaluates an exact compiled
contract with a dense factorization, not the colored oracle path of
Section~\ref{sec:gmrf}, so the measured cost ratios do not carry the coloring
penalty that the refusal is about; and at $K=64$ four observations per scale
leave the core weakly identified and close to prior-dominated, which is visible
in the reference chains rather than hidden by them.

\FloatBarrier
\subsection{Discovery census over PosteriorDB}
\label{app:census}

We apply the same discovery procedure to the 46 targets of
Section~\ref{sec:census} (45 PosteriorDB posteriors with shipped references
and the DBN), plus pilots, radon, rats, and uk\_drivers, whose references are
constructed in the supplement. All 50 use three random-probe restarts and
default tolerances. Table~\ref{tab:census} includes unsuccessful candidates
and reports dimensions, precision structure, probe colors, held-out residuals,
and discovery time.

\begin{table}[p]
 \centering
 \footnotesize
 \renewcommand{\arraystretch}{0.84}
 \setlength{\tabcolsep}{4pt}
 \caption{Discovery on 50 models. ``Declined'' rows have no
 conditionally Gaussian block and leave the sampler unchanged;
 ``structure'' describes the numerically validated precision pattern.
 Entries use the largest candidate found, before cost-aware refinement
 (radon sampling uses a $170/5$ split). $^{a}$Probe modes find 24 and 25
 coordinates; the automatic result is shown. $^{b}$The 49-coordinate random
 candidate fails validation at $1.1\times10^{-8}$; plausible probes validate
 the reported 40-coordinate block. Model names are abbreviated.}
 \label{tab:census}
 \fittable{\begin{tabular}{@{}lrrrrlcr@{}}
\toprule
model & $D$ & $d_g$ & $d_h$ & colors & structure & identity residual & disc.\ (s)\\
\midrule
\textsf{arK-arK} & 7 & 6 & 1 & 6 & dense & $1{\cdot}10^{-13}$ & 0.06 \\
\textsf{arma-arma11} & 4 & 1 & 3 & 1 & independent & $2{\cdot}10^{-14}$ & 0.18 \\
\textsf{bball\_drive\_0-hmm\_drive\_0} & 6 & -- & 6 & -- & none & \emph{declined} & 0.55 \\
\textsf{bball\_drive\_1-hmm\_drive\_1} & 6 & -- & 6 & -- & none & \emph{declined} & 0.54 \\
\textsf{dbn\_lgss-dbn\_lgss} & 208 & 200 & 8 & 6 & banded & $2{\cdot}10^{-14}$ & 0.22 \\
\textsf{diamonds-diamonds}$^{a}$ & 26 & 24 & 2 & 24 & dense & $2{\cdot}10^{-12}$ & 0.10 \\
\textsf{earnings-earn\_height} & 3 & 2 & 1 & 2 & dense & $1{\cdot}10^{-11}$ & 0.01 \\
\textsf{earnings-log10earn\_height} & 3 & 2 & 1 & 2 & dense & $1{\cdot}10^{-11}$ & 0.01 \\
\textsf{earnings-logearn\_height} & 3 & 2 & 1 & 2 & dense & $1{\cdot}10^{-12}$ & 0.01 \\
\textsf{earnings-logearn\_height\_male} & 4 & 3 & 1 & 3 & dense & $4{\cdot}10^{-12}$ & 0.03 \\
\textsf{earnings-logearn\_interaction} & 5 & 4 & 1 & 4 & dense & $1{\cdot}10^{-12}$ & 0.03 \\
\textsf{earnings-logearn\_interaction\_z} & 5 & 4 & 1 & 4 & dense & $9{\cdot}10^{-13}$ & 0.01 \\
\textsf{earnings-logearn\_loght\_male} & 4 & 3 & 1 & 3 & dense & $9{\cdot}10^{-14}$ & 0.01 \\
\textsf{eight\_schools-centered} & 10 & 9 & 1 & 9 & dense & $2{\cdot}10^{-14}$ & 0.01 \\
\textsf{eight\_schools-noncentered} & 10 & 9 & 1 & 9 & dense & $2{\cdot}10^{-16}$ & 0.01 \\
\textsf{garch-garch11} & 4 & -- & 4 & -- & none & \emph{declined} & 0.52 \\
\textsf{gp\_pois\_regr-gp\_pois\_regr} & 13 & -- & 13 & -- & none & \emph{declined} & 0.03 \\
\textsf{gp\_pois\_regr-gp\_regr} & 3 & -- & 3 & -- & none & \emph{declined} & 0.03 \\
\textsf{hmm\_example-hmm\_example} & 4 & -- & 4 & -- & none & \emph{declined} & 0.13 \\
\textsf{kidiq-kidscore\_interaction} & 5 & 4 & 1 & 4 & dense & $4{\cdot}10^{-14}$ & 0.02 \\
\textsf{kidiq-kidscore\_momhs} & 3 & 2 & 1 & 2 & dense & $2{\cdot}10^{-13}$ & 0.01 \\
\textsf{kidiq-kidscore\_momhsiq} & 4 & 3 & 1 & 3 & dense & $6{\cdot}10^{-14}$ & 0.01 \\
\textsf{kidiq-kidscore\_momiq} & 3 & 2 & 1 & 2 & dense & $4{\cdot}10^{-14}$ & 0.01 \\
\textsf{kidiq\_mw-interaction\_c} & 5 & 4 & 1 & 4 & dense & $3{\cdot}10^{-13}$ & 0.02 \\
\textsf{kidiq\_mw-interaction\_c2} & 5 & 4 & 1 & 4 & dense & $2{\cdot}10^{-13}$ & 0.01 \\
\textsf{kidiq\_mw-interaction\_z} & 5 & 4 & 1 & 4 & dense & $2{\cdot}10^{-13}$ & 0.01 \\
\textsf{kidiq\_mw-kidscore\_mom\_work} & 5 & 4 & 1 & 4 & dense & $1{\cdot}10^{-13}$ & 0.02 \\
\textsf{kilpisjarvi\_mod-kilpisjarvi} & 3 & 2 & 1 & 2 & dense & $2{\cdot}10^{-9}$ & 0.01 \\
\textsf{low\_dim\_gauss\_mix} & 5 & -- & 5 & -- & none & \emph{declined} & 0.03 \\
\textsf{mcycle\_gp-accel\_gp}$^{b}$ & 66 & 40 & 26 & 40 & dense & $7{\cdot}10^{-14}$ & 0.07 \\
\textsf{mesquite-logmesquite} & 8 & 7 & 1 & 7 & dense & $9{\cdot}10^{-13}$ & 0.01 \\
\textsf{mesquite-logmesquite\_logva} & 5 & 4 & 1 & 4 & dense & $4{\cdot}10^{-13}$ & 0.01 \\
\textsf{mesquite-logmesquite\_logvas} & 8 & 7 & 1 & 7 & dense & $6{\cdot}10^{-13}$ & 0.01 \\
\textsf{mesquite-logmesquite\_logvash} & 7 & 6 & 1 & 6 & dense & $7{\cdot}10^{-13}$ & 0.01 \\
\textsf{mesquite-logmesquite\_logvolume} & 3 & 2 & 1 & 2 & dense & $3{\cdot}10^{-13}$ & 0.01 \\
\textsf{mesquite-mesquite} & 8 & 7 & 1 & 7 & dense & $4{\cdot}10^{-13}$ & 0.01 \\
\textsf{nes1972-nes} & 10 & 9 & 1 & 9 & dense & $2{\cdot}10^{-14}$ & 0.06 \\
\textsf{nes1976-nes} & 10 & 9 & 1 & 9 & dense & $3{\cdot}10^{-14}$ & 0.03 \\
\textsf{nes1980-nes} & 10 & 9 & 1 & 9 & dense & $9{\cdot}10^{-15}$ & 0.06 \\
\textsf{nes1984-nes} & 10 & 9 & 1 & 9 & dense & $6{\cdot}10^{-15}$ & 0.02 \\
\textsf{nes1988-nes} & 10 & 9 & 1 & 9 & dense & $1{\cdot}10^{-14}$ & 0.02 \\
\textsf{nes1992-nes} & 10 & 9 & 1 & 9 & dense & $3{\cdot}10^{-14}$ & 0.04 \\
\textsf{nes1996-nes} & 10 & 9 & 1 & 9 & dense & $2{\cdot}10^{-14}$ & 0.03 \\
\textsf{nes2000-nes} & 10 & 9 & 1 & 9 & dense & $6{\cdot}10^{-15}$ & 0.01 \\
\textsf{sblrc-blr} & 6 & 5 & 1 & 5 & dense & $2{\cdot}10^{-11}$ & 0.01 \\
\textsf{sblri-blr} & 6 & 5 & 1 & 5 & dense & $3{\cdot}10^{-11}$ & 0.01 \\
\midrule
\textsf{pilots-pilots} & 18 & 15 & 3 & 14 & dense & $6{\cdot}10^{-11}$ & 0.05 \\
\textsf{radon\_mn-var\_int\_slope\_noncent} & 175 & 172 & 3 & 147 & dense & $4{\cdot}10^{-15}$ & 0.13 \\
\textsf{rats\_data-rats\_model} & 65 & 62 & 3 & 31 & dense & $1{\cdot}10^{-12}$ & 0.02 \\
\textsf{uk\_drivers\_unb-stoch\_level\_seas} & 389 & 386 & 3 & 386 & dense & $5{\cdot}10^{-9}$ & 0.23 \\
\bottomrule
\end{tabular}
}
\end{table}

The procedure returns 42 numerical certificates, with relative identity
residuals between $2\times10^{-16}$ and $5\times10^{-9}$ and blocks ranging
from one coordinate (arma) to 386 of 389 (uk\_drivers), and declines seven
targets.
None of the seven declined targets has an exactly Gaussian conditional block.
The remaining target, mcycle\_gp, has a 49-coordinate random-probe candidate
rejected at the held-out pattern check; probes at plausible locations validate
a 40-coordinate block. Probe stability and sensitivity are reported in
Appendix~\ref{app:discovery}. Total discovery time is 3.42 seconds, followed
by 0.4 seconds for construction and validation on ten CPU cores; the original
46-target subset accounts for 3.01 and 0.26 seconds. Checking for a numerical
certificate is therefore inexpensive relative to any sampling run. These runs
were made on the same machine as the main experiments, outside the container.
The only required input is a differentiable log density: discovery, numerical
validation, and target-preserving map construction are automatic, and the
resulting metric can be passed to any sampler that accepts one. Removing the
residual additionally requires an exact contract.

\paragraph{Sampling on real-data targets.}
Table~\ref{tab:wildlayer} compares raw and transformed targets on pilots,
radon, and rats using Fisher-diagonal NUTS, eight chains, 500 retained draws,
and five seeds (three for the expensive radon layer). The transformed runs
use 150 warmup iterations, compared with 400 for the raw target. Reducing raw
pilots warmup to 150 does not resolve its sampling failures: the median has
397 divergences, compared with 450 at 400 warmup.

The raw radon and rats samplers are already divergence-free. The transformed
runs agree with their references and retain zero divergences, but the generic
dense adapter is about two orders of magnitude slower on radon. A follow-up
separates this cost from the choice of representation. Moving the two global
means into the core reduces the precision recovery from 147 to two colors.
Generic marginalization still requires eight oracle passes per score;
using a fixed CONGA metric instead needs one target evaluation per NUTS step.
Its median sampling time is 40.1 seconds, or 70.3 including construction,
compared with 55.9 seconds raw. The preliminary core fit has split
$\rhat=1.07\ [1.03,1.36]$, so its quality remains a routing diagnostic.
For this example, preconditioning helps only when construction can be reused;
the original target is cheaper for a single short run.

\begin{table}[t]
 \centering
 \footnotesize
 \setlength{\tabcolsep}{3.5pt}
 \caption{Raw target versus the generic certified layer on three real-data targets,
 median with min--max bands over five seeds (three for the radon layer arm).
 Mean-$z$ error is the coordinate-average absolute mean difference in reference
 standard deviations and scale ratio the coordinate-average sampled-to-reference
 marginal SD ratio; the parenthesis after each target is $D/d_g$.
 $^{\dagger}$wall clocks spanning a host suspension are excluded from the
 time band.}
 \label{tab:wildlayer}
 \fittable{\begin{tabular}{@{}llcccc@{}}
\toprule
target & arm & mean-$z$ error & scale ratio & divergences & time (s)\\
\midrule
\textsf{pilots} (18/15) & raw w400 & $0.035\ [0.023,0.324]$ & $0.99\ [0.86,1.14]$ & $450\ [394,671]$ & $255\ [245,270]$ \\  %
 & certified layer w150 & $0.021\ [0.010,0.040]$ & $1.01\ [0.98,1.03]$ & $0\ [0,0]$ & $92\ [86,117]$ \\  %
\addlinespace
\textsf{radon\_mn} (175/170) & raw w400 & $0.014\ [0.014,0.015]$ & $1.00\ [1.00,1.00]$ & $0\ [0,0]$ & $59\ [52,73]$ \\  %
 & certified layer w150 & $0.017\ [0.014,0.018]$ & $1.00\ [1.00,1.00]$ & $0\ [0,0]$ & $4442\ [1285,7598]$$^{\dagger}$ \\  %
\addlinespace
\textsf{rats} (65/60) & raw w400 & $0.014\ [0.013,0.015]$ & $1.00\ [1.00,1.00]$ & $0\ [0,0]$ & $23\ [20,28]$ \\  %
 & certified layer w150 & $0.015\ [0.013,0.015]$ & $1.00\ [1.00,1.01]$ & $0\ [0,0]$ & $119\ [114,183]$ \\  %
\bottomrule
\end{tabular}
}
\end{table}

\subsection{Stock--Watson conditional-path case study}
\label{app:stock-watson}

The Stock--Watson model \cite{stockwatson2007} represents quarterly inflation
$y_t$ using persistent log volatility $z_t$, transient log volatility $x_t$
and a latent level $\tau_t$:
\begin{align}
 z_t\mid z_{t-1},\sigma
   &\sim \N(z_{t-1},\sigma^2), &&t=2,\ldots,T-1,\\
 x_t\mid x_{t-1},\sigma
   &\sim \N(x_{t-1},\sigma^2), &&t=2,\ldots,T,\\
 \tau_t\mid\tau_{t-1},z_{t-1}
   &\sim \N(\tau_{t-1},e^{z_{t-1}}), &&t=2,\ldots,T,\\
 y_t\mid\tau_t,x_t
   &\sim \N(\tau_t,e^{x_t}), &&t=1,\ldots,T,
 \label{eq:stock-watson}
\end{align}
with $\sigma^{-2}\sim\operatorname{Gamma}(5,0.5)$ in the rate
parameterization. We use the public $T=252$ quarterly series and the innovation
coordinates for all three latent paths, with $\ell=\log\sigma^2$. The raw
target therefore has $3T=756$ coordinates.

For compilation, CONGA exposes $\tau_{1:T}$ in centered coordinates. Conditional
on $(z,x)$ it is exactly Gaussian with tridiagonal precision. Writing
$r_t=e^{-z_t}$ and $o_t=e^{-x_t}$, its off-diagonal entries are
$\Lambda_{t,t+1}=-r_t$, its diagonal entries are
\begin{equation}
 \Lambda_{11}=o_1+r_1,\qquad
 \Lambda_{tt}=o_t+r_{t-1}+r_t\ (2\le t<T),\qquad
 \Lambda_{TT}=o_T+r_{T-1},
 \label{eq:stock-watson-precision}
\end{equation}
and its canonical shift is $b_t=o_ty_t$. Exact tridiagonal elimination
integrates out the 252-dimensional path; conditional reconstruction uses
$\N(\Lambda^{-1}b,\Lambda^{-1})$. The contract has $d_g=252$, $d_h=504$,
three distance-two colors, half-bandwidth one and 754 structural nonzeros.
Independent PyTorch and JAX implementations agree to absolute error
$4.6\times10^{-13}$ in the raw value, $3.2\times10^{-12}$ in the raw gradient,
$4.3\times10^{-14}$ in the collapsed value and $1.2\times10^{-13}$ in the
collapsed gradient; the checked joint conditional identity has residual
$1.2\times10^{-13}$.

\paragraph{Protocol.}
All four arms use float64 JAX-compiled value-and-gradient callbacks; NUTS and
WALNUTS share the same guarded callback within each parameterization, and no
hand-derived score is used. NUTS uses nutpie's diagonal adaptation with
target acceptance $0.95$ and maximum tree depth ten. WALNUTS uses the released
adaptive implementation initialized at the Stock--Watson paper's macro step
$h=0.1$, with local energy threshold $\delta=0.3$ and a minimum of eight
microsteps per macrostep. Each seed runs eight chains for 800 warmup and 800
retained iterations. Initial core coordinates are zero in both parameterizations;
the raw $\tau$ path starts at its conditional mean at that core point.
Collapsed draws reconstruct $\tau$ exactly with a fixed independent Gaussian
stream, and diagnostics are computed in the common $(z,x,\tau,\ell)$
coordinates. Score calls include warmup and sampling. Reported seconds include
both phases but exclude one-time JAX compilation (at most 0.20 seconds) and
common-coordinate reconstruction (at most 0.20 seconds).

The shared score maps any non-finite value or gradient arising outside the
stable floating-point domain to $-10^{30}$ with a zero gradient. This guard is
used identically by NUTS and WALNUTS within each parameterization: the raw and collapsed WALNUTS arms trigger it 852
and 33 times, respectively, over $9.18$ and $3.37$ million score calls.

\begin{figure}[t]
 \centering
 \includegraphics[width=\linewidth]{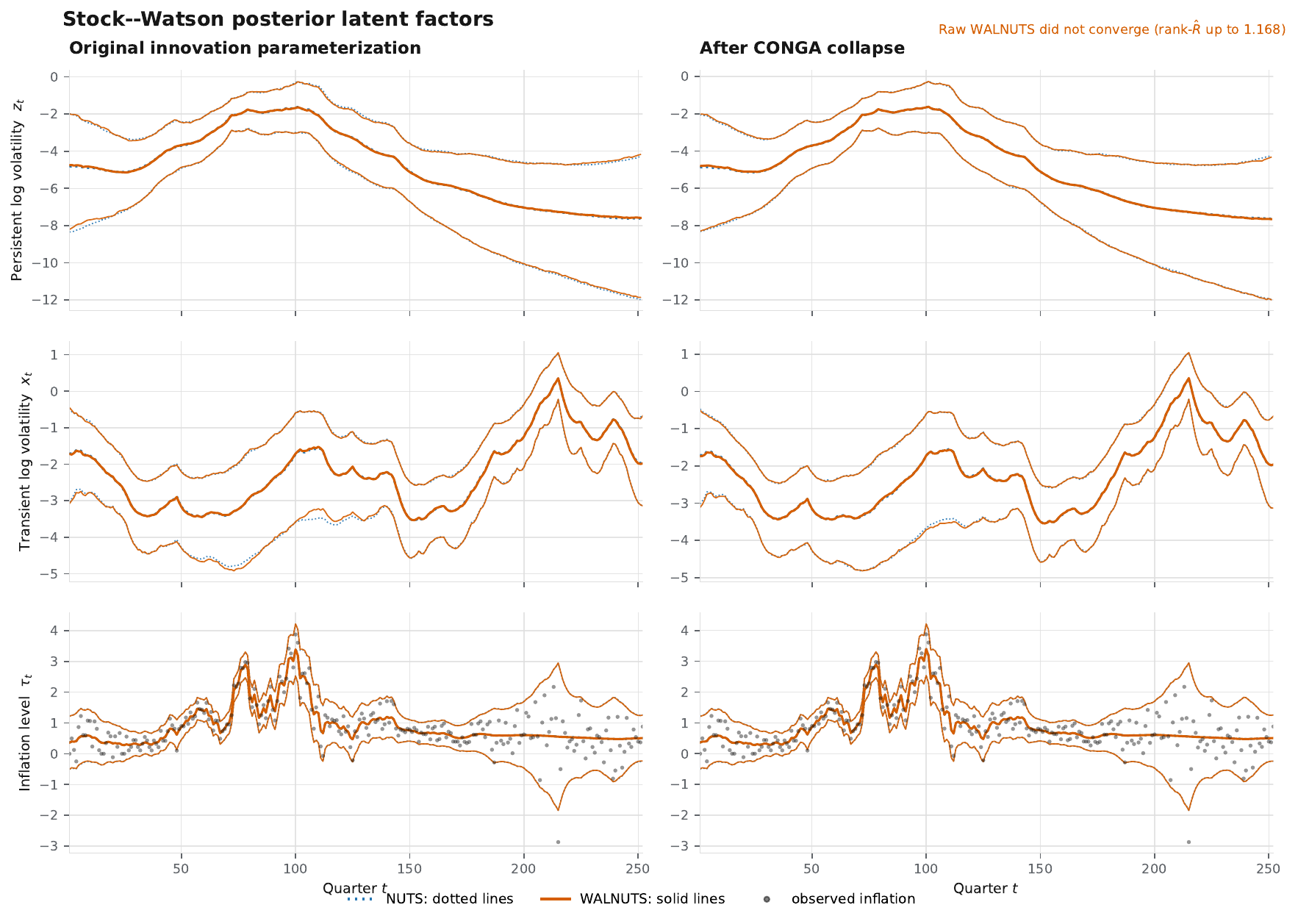}
 \caption{Stock--Watson latent factors, raw (left) and collapsed with exact
 reconstruction (right), pooled over three seeds of eight chains and 800
 draws. Curves show the 0.05, 0.50, and 0.95 quantiles for $z_t,x_t$; for
 $\tau_t$, they show its median and the pointwise medians of
 $\tau_t\pm2\exp(x_t/2)$, as in the WALNUTS paper. Dots are observations.
 NUTS is dotted blue; WALNUTS is solid orange. Raw WALNUTS fails convergence
 despite the visual agreement (Figure~\ref{fig:stock-watson-diagnostics}).}
 \label{fig:stock-watson-factors}
\end{figure}

\begin{figure}[t]
 \centering
 \includegraphics[width=\linewidth]{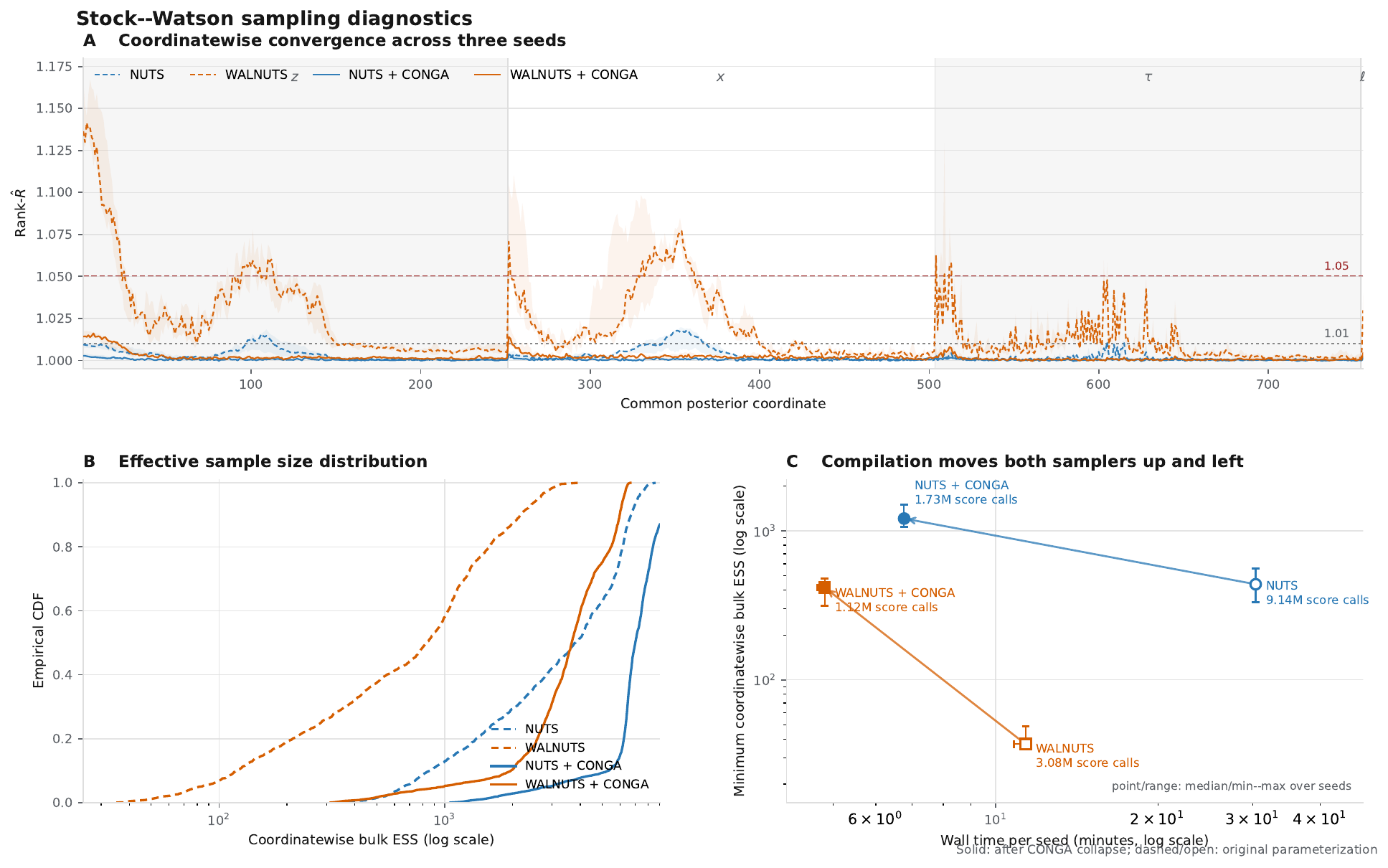}
 \caption{Coordinatewise diagnostics and cost for the matched Stock--Watson
 experiment. Panel A plots the median rank-$\rhat$ profile over seeds, with the
 min--max seed envelope; the raw WALNUTS failure spans the $z$, $x$ and $\tau$
 paths. Panel B gives the empirical CDF of coordinatewise bulk ESS over all
 three seeds. Panel C compares median wall time and minimum bulk ESS, with
 min--max seed ranges and median score calls annotated; arrows connect each raw
 arm to its CONGA counterpart. Solid curves and filled points denote CONGA,
 while dashed curves and open points denote the original parameterization.}
 \label{fig:stock-watson-diagnostics}
\end{figure}

Figure~\ref{fig:stock-watson-factors} compares the latent-path summaries.
Despite their close agreement, raw WALNUTS fails the coordinatewise
convergence checks in Figure~\ref{fig:stock-watson-diagnostics}.

\begin{table}[ht]
 \centering
 \footnotesize
 \setlength{\tabcolsep}{3.2pt}
\caption{Matched Stock--Watson sampling, three seeds. Entries are medians with
min--max seed ranges in brackets; divergences are summed over seeds during the
retained phase and are not exposed by WALNUTS. Guard hits are summed over
seeds. Minimum bulk ESS and maximum rank-$\rhat$ range over all 756 common-space
coordinates after exact reconstruction.}
\label{tab:stock-watson-four-arm}
\fittable{%
 \begin{tabular}{@{}lrrrrrrr@{}}
  \toprule
  arm & dim. & seconds & score calls ($10^6$) & min. bulk ESS
      & max. rank-$\rhat$ & divergences & guard hits \\
  \midrule
  NUTS raw
    & 756 & 1820.5 [1816.5, 1848.5] & 9.141 [9.022, 9.198]
    & 437 [330, 560] & 1.018 [1.011, 1.020] & 64 & 0 \\
  WALNUTS raw
    & 756 & 682.1 [648.8, 690.4] & 3.082 [2.959, 3.141]
    & 37 [35, 49] & 1.167 [1.138, 1.168] & -- & 852 \\
  NUTS + CONGA
    & 504 & 406.2 [401.0, 409.8] & 1.731 [1.721, 1.749]
    & 1207 [1062, 1492] & 1.004 [1.003, 1.004] & 4 & 4 \\
  WALNUTS + CONGA
    & 504 & 289.3 [280.7, 290.9] & 1.117 [1.116, 1.136]
    & 416 [312, 476] & 1.018 [1.016, 1.022] & -- & 33 \\
  \bottomrule
 \end{tabular}}
\end{table}

Both NUTS parameterizations pass the $\rhat\le1.05$ criterion, and collapse
improves their minimum ESS per second by $12.5\times$. This is the median of
matched-seed ratios with reconstruction included, so it differs from the
ratio of the medians in Table~\ref{tab:stock-watson-four-arm}. Raw WALNUTS
fails the criterion in every seed and its collapsed version passes. The
corresponding nominal ratio is $23.7\times$, but the ESS of an unconverged
chain is a failure diagnostic, not an efficiency estimate. On the collapsed
target, where both samplers pass, WALNUTS is faster per run, while NUTS has
$2.16\times$ higher minimum ESS per second.
Collapsed NUTS retains four sampling divergences across the three seeds.
Thus the local integrator adaptation and structural compilation are
complementary, with CONGA providing the decisive mixing repair for the raw
WALNUTS parameterization.

\FloatBarrier

\subsection{Electricity-load DLM: a temporal path of spatial graphs}
\label{app:dlm-load}

We combine standard DLM components over a transmission grid to obtain a
model with temporal and spatial dependence: hourly load $y_{r,t}$ in zones
$r=1,\ldots,R$ over days $d(t)$ obeys
\begin{equation}
 y_{r,t}=\mu_{r,t}+s^{d}_{r,t}+s^{w}_{r,t}
 +\mathbf b_{r,t}^\top\boldsymbol\phi(\mathrm{temp}_{r,t})
 +h_{r,t}I^{\mathrm{hol}}_t+\lambda_rf_t+v_{r,t},
 \qquad v_{r,t}\sim\N\!\bigl(0,e^{x_{d(t)}+c_r}\bigr),
 \label{eq:dlm-obs}
\end{equation}
with a local linear trend, harmonic seasonal blocks, a mean-reverting
temperature response on a spline basis $\boldsymbol\phi$, a holiday random
walk and a national AR(1) factor:
\begin{align}
 \mu_{r,t}&=\mu_{r,t-1}+\beta_{r,t-1}+w^{\mu}_{r,t}, &
 \mathbf w^{\mu}_t&\sim\N\!\bigl(\mathbf 0,
   e^{z_{d(t)}}(Q_{\mathcal G}+\kappa I)^{-1}\bigr),\nonumber\\
 \begin{pmatrix}a_{j,t}\\b_{j,t}\end{pmatrix}
 &=\begin{pmatrix}\cos\omega_{cj}&\sin\omega_{cj}\\
    -\sin\omega_{cj}&\cos\omega_{cj}\end{pmatrix}
   \begin{pmatrix}a_{j,t-1}\\b_{j,t-1}\end{pmatrix}+\mathbf w^{c}_{j,t}, &
 s^{c}_{r,t}&=\textstyle\sum_ja_{r,c,j,t},\nonumber\\
 \mathbf b_{r,t}&=\rho\,\mathbf b_{r,t-1}+(1-\rho)\bar{\mathbf b}_r
   +\mathbf w^{b}_{r,t}, &
 f_t&=\varphi_ff_{t-1}+w^{f}_t,\quad
 w^{f}_t\sim\N\!\bigl(0,e^{z^{f}_{d(t)}}\bigr),
 \label{eq:dlm-states}
\end{align}
where $Q_{\mathcal G}$ is the transmission-graph Laplacian, $\beta$ and $h$
are Gaussian random walks, and the long-run responses $\bar{\mathbf b}_r$ are
static Gaussian coordinates. The core contains the coefficients and scales that make the joint model
non-Gaussian: the three daily log-volatility random
walks $z,z^{f},x$ share one vol-of-vol $\sigma_v$ exactly as in
Appendix~\ref{app:stock-watson}, joined by $\kappa$, $\rho$, $\varphi_f$, the
loadings $\lambda_{2:R}$ ($\lambda_1=1$; they multiply $f_t$ bilinearly), the
noise offsets $c_{2:R}$ ($c_1=0$) and per-component innovation
log-variances, all with Gaussian priors on unconstrained coordinates. Data
are simulated: diurnal temperatures with autoregressive weather, a two-day
heat wave in two adjacent zones with a coincident trend-volatility spike, and
one holiday. We use these simulated data to assess the inference procedure.

At the matched-inference scale ($R=3$ zones on a path graph, $T=168$ hours,
three daily and one weekly harmonic, two spline arms) each hour carries 40
dynamic states, so the fiber has $168\times40+6=6{,}726$ coordinates over a
34-dimensional core. Model algebra supplies the exact contract: conditional
on the core the trajectory is jointly Gaussian, the collapsed score is the
sequential Kalman prediction-error decomposition, and reconstruction is a
backward sampling pass over the non-redundant coordinates (full state at
$t=T$, dynamic subvector conditionals for $t<T$, the statics carried as
copies). The direct joint density, collapsed filter, and backward conditional are
implemented independently. Their conditional-identity residual is at most
$2.4\times10^{-11}$ relative and varies across $x$ at fixed $\eta$ by at
most $6\times10^{-8}$ nats. The recovered precision is constant in $x$ and
block-tridiagonal in time, apart from the static-coordinate connections.
Time-varying temperature and holiday terms make it non-Toeplitz. It also fails
to commute across core points (relative commutator $5.5\times10^{-3}$,
compared with $1.9\times10^{-18}$ for the GMRF pencil). A shared spectral
basis is therefore unavailable. CONGA uses sequential temporal elimination
with an $R\times R$ graph solve in each block.

\paragraph{Protocol.}
Both arms use guarded float64 JAX value-and-gradient callbacks behind
nutpie's diagonal adaptation with target acceptance $0.90$ and maximum tree
depth ten; each seed runs four chains for 600 warmup and 600 retained
iterations, seeds 0--2. Initial core coordinates are zero in both
parameterizations; the raw fiber starts at its conditional mean at that core
point. Collapsed draws reconstruct all 6{,}726 fiber coordinates exactly
with a fixed independent Gaussian stream, and diagnostics range over all
6{,}760 common coordinates. Score latencies are $0.10$ ms (raw) and $22.2$
ms (collapsed), a $218\times$ premium per gradient; compilation takes at
most $0.8$ seconds. Reported seconds cover warmup and sampling;
reconstruction adds at most $36.2$ seconds and is included in the
ESS-per-second ratios below. The guard never fires in either arm
($14.7$ million raw and $0.23$ million collapsed calls).

\begin{table}[ht]
 \centering
 \footnotesize
 \setlength{\tabcolsep}{3.2pt}
\caption{Matched load-DLM sampling, three seeds. Entries are medians with
min--max seed ranges in brackets; divergences and guard hits are summed over
seeds. Minimum bulk ESS and maximum rank-$\rhat$ range over all 6{,}760
common-space coordinates after exact reconstruction.}
\label{tab:dlm-load}
\fittable{%
 \begin{tabular}{@{}lrrrrrrr@{}}
  \toprule
  arm & dim. & seconds & score calls ($10^6$) & min. bulk ESS
      & max. rank-$\rhat$ & divergences & guard hits \\
  \midrule
  NUTS raw
    & 6{,}760 & 630.5 [624.6, 653.5] & 4.894 [4.865, 4.894]
    & 5.3 [5.3, 5.6] & 2.113 [1.910, 2.121] & 30 & 0 \\
  NUTS + CONGA
    & 34 & 566.1 [543.5, 574.0] & 0.077 [0.075, 0.078]
    & 1161 [1148, 1240] & 1.005 [1.005, 1.006] & 3 & 0 \\
  \bottomrule
 \end{tabular}}
\end{table}

Table~\ref{tab:dlm-load} shows a categorical split rather than a speedup.
Every raw seed fails the rank-$\rhat\le1.05$ gate at $1.91$--$2.12$ with
minimum bulk ESS near five, spending essentially every iteration at the
1{,}023-step depth budget: the three volatility funnels and the bilinear
$\lambda_rf_t$ ridge over a 6{,}726-dimensional centered fiber defeat NUTS
outright. Every collapsed seed passes at worst rank-$\rhat=1.006$ with
minimum bulk ESS at least 1{,}148 on 15-step trees, in comparable wall time.
The nominal ratio of median minimum bulk ESS per second, including
reconstruction, is $234\times$; because the raw chains do not converge, we
read it as a failure diagnostic rather than a speedup. The collapsed
posterior covers all eleven spot-checked core
truths ($\kappa$, $\rho$, $\varphi_f$, $\sigma_v$, both loadings, both noise
offsets and one point per volatility path) in their central 90\% intervals.

We also evaluate a larger model with $R=14$ zones on a meshed 20-edge grid,
$T=8{,}760$ hours, eight daily, four weekly, and three annual harmonics, and
four spline arms. It has 519 dynamic states per hour, a 575-dimensional
Kalman state including statics, and a $4{,}546{,}496$-coordinate fiber over
a 1{,}131-dimensional core. Storing the current discovery procedure's dense
conflict graph would require terabytes, so we supply the algebraic contract.
A collapsed evaluation takes 50.6 seconds and a value-and-gradient call with
per-day rematerialization takes 213.7 seconds on a laptop CPU. Gradients are
finite and values agree exactly between rematerialized and plain schedules.
Posterior sampling at this scale was not attempted.

\FloatBarrier

\subsection{A misspecified certificate: Student-t sensor noise}
\label{app:misspec}

Section~\ref{sec:certificate} separates numerical evidence from sampler
validity: a wrong conditional-Gaussian identity costs efficiency, never
correctness, as long as the
sampler targets the exact transformed density of Eq.~\eqref{eq:failsafe}.
This appendix exercises that separation end to end on the DBN of
Section~\ref{sec:dbn}.

\paragraph{Setup.}
One dataset ($T=100$, 208 coordinates) is simulated from the DBN of
Section~\ref{sec:experiments} with Student-t$(4)$ noise on the dense $y$
sensor, giving three innovations beyond
$3\sigma_y$, the largest at $6.8\sigma_y$. A model ladder
$\nu\in\{\infty,20,4,2.5\}$ is then fitted to this fixed dataset, where
$\nu=\infty$ is the Gaussian model of Section~\ref{sec:dbn} and $\nu=4$
matches the simulation. A conditional-Gaussian candidate map is discovered and
constructed once from the \emph{Gaussian} model's value-and-gradient oracle; on
every finite-$\nu$ rung its identity is wrong by construction. This deliberately
models a practitioner who retains an old transport after robustifying the
likelihood, while continuing to evaluate the new target exactly.

\paragraph{Reference posteriors.}
Student-t noise is an inverse-gamma scale mixture of Gaussians, so
augmenting one mixing variable $\lambda_t$ per $y$ observation makes the
model conditionally linear-Gaussian again. Each rung's reference is drawn
by a partially collapsed Gibbs sampler with exact kernels: $\eta$ given
$\lambda$ by Metropolis--Hastings on the $\lambda$-conditioned
Kalman-collapsed posterior (states integrated out analytically, walker-%
difference proposals over a 64-walker ensemble), states given
$(\eta,\lambda)$ by forward-filter backward-sampling, and $\lambda$ given
the rest conjugate. No arm being evaluated contributes to its own
reference. Each rung retains 16{,}000 draws; the worst split $\rhat$ over
the ensemble, across all four rungs, is 1.033 on the core coordinates and
1.007 on the states. At $\nu=\infty$, an independent IS + DE-MC calculation
agrees on every
hyperparameter mean within 0.019 reference standard deviations. That
calculation reaches only split $\rhat=1.11$, so convergence assessment rests
on the collapsed Gibbs reference. The $\nu=\infty$ and $\nu=2.5$ references
required two to three times as many ensemble iterations as the intermediate
cases.

\paragraph{Discovery under Student-$t$ noise.}
Rerunning discovery on the Student-t oracle itself does not reproduce the
200-state candidate: at every finite $\nu$ the probes detect the
non-affine score of the $u$ chain (the coordinates the $y$ sensor touches)
and remove all 100 of them, returning a 102-coordinate block --- the $v$
chain plus the two coupling coefficients $a_{11},a_{21}$, which multiply
only core coordinates and therefore remain exactly jointly Gaussian with
$v$. Discovery thus finds the correct smaller candidate on its own. The
sampling experiment below deliberately retains the full-state map
from the Gaussian model to assess the effect of misspecification.

\begin{table}[t]
 \centering
 \footnotesize
 \setlength{\tabcolsep}{4.5pt}
 \caption{Certificate diagnostics per rung. ``disc.\ $d_g$'' is the block
 discovery returns when probing the Student-t oracle directly. The residual
 columns evaluate Eq.~\eqref{eq:residual} for the \emph{forced} Gaussian
 certificate at 256 reference draws; $\hat\eps$ is the largest observed
 $|r|$. ``Accept.'' is the measured acceptance of the corrected conditional
 independence proposal of Proposition~\ref{prop:robustness}, to be compared
 with its worst-case bound $e^{-2\hat\eps}$.}
 \label{tab:misspec-cert}
 \fittable{\begin{tabular}{@{}lrccccc@{}}
\toprule
model $\nu$ & disc.\ $d_g$ & $|r|$ median & $|r|$ q95 & $\hat\eps$ & accept.\ measured & bound $e^{-2\hat\eps}$\\
\midrule
$\infty$ (exact) & 200 & $3{\cdot}10^{-14}$ & $8{\cdot}10^{-13}$ & $2{\cdot}10^{-10}$ & 1.000 & $1$ \\
20 & 102 & $1.44$ & $5.9$ & $10.5$ & 0.358 & $8{\cdot}10^{-10}$ \\
4 & 102 & $10.1$ & $50.4$ & $136$ & 0.010 & $8{\cdot}10^{-119}$ \\
2.5 & 102 & $17.2$ & $128$ & $1{\cdot}10^{3}$ & 0.000 & 0 \\
\bottomrule
\end{tabular}
}
\end{table}

\paragraph{Residuals and the acceptance bound.}
Table~\ref{tab:misspec-cert} quantifies how wrong the forced certificate
is. At $\nu=\infty$ the identity holds to numerical precision and the
corrected independence proposal accepts every draw, matching the bound
exactly. At finite $\nu$ the residual grows to tens or hundreds of nats
--- the glitch observations dominate $|r|$ --- and measured acceptance
decays from $0.36$ ($\nu=20$) to essentially zero ($\nu=2.5$). The
$e^{-2\eps}$ bound is never violated but is loose: it is a worst-case
statement over the tested region, and once $\hat\eps$ reaches tens of nats
it is vacuous. The practical reading is that independence-style corrections
of a badly wrong certificate stall, which is why the fail-safe uses the
transformed density inside a gradient-based sampler instead.

\paragraph{Sampling through the misspecified map.}
We compare Fisher-diagonal NUTS on the raw target (400 warmup iterations) and
on Eq.~\eqref{eq:failsafe} (150 warmup iterations), with four chains, 500
draws, and three seeds per noise model. The generic dense adapter requires
$2+6$ oracle passes per evaluation. An uncorrected control uses draws from
the Gaussian-model posterior as if the false factorization were valid.

The fail-safe arm is statistically indistinguishable from the reference at
every rung: errors are 0.015--0.019 reference standard deviations,
versus 0.016--0.017 when subsampling the references themselves. They pass the
convergence criterion at every $\nu$, although minimum ESS falls as the
likelihood departs from Gaussianity. Correctness does not depend on the
certificate being right. The uncorrected control instead develops
substantial mean bias. Raw NUTS fails the convergence criterion at every
$\nu$, so its error estimates in Table~\ref{tab:misspec-sampling} are
diagnostic. Even a badly wrong certificate, executed through the exact
transformed density, is here more reliable than no transport at all. The
study took about five hours, plus 90 minutes for discarded pilots, outside
the container.

\begin{table}[htbp]
 \centering
 \footnotesize
 \setlength{\tabcolsep}{3pt}
 \caption{Sampling under a misspecified certificate, medians and min--max
 ranges over three seeds. ``Raw'' and ``layer'' denote NUTS on the original
 and exactly transformed targets; ``uncorrected'' uses the false Gaussian
 factorization. Errors are measured against independent references.
 Divergences count retained sampling; ``grads'' includes warmup. Each layer
 evaluation requires $2+6$ oracle passes.}
 \label{tab:misspec-sampling}
 \fittable{\begin{tabular}{@{}llcccccc@{}}
\toprule
$\nu$ & arm & mean-$z$ error & scale ratio & div.\ & max $\rhat$ & min ESS & grads ($10^3$)\\
\midrule
$\infty$ (exact) & raw & $0.062\ [0.024,0.068]$ & $0.99\ [0.99,1.01]$ & $3\ [0,4]$ & $1.32\ [1.08,1.63]$ & $11\ [7,43]$ & $211\ [209,338]$ \\
 & layer & $0.016\ [0.016,0.018]$ & $1.00\ [1.00,1.00]$ & $1\ [0,2]$ & $1.01\ [1.01,1.01]$ & $686\ [611,747]$ & $142\ [51,164]$ \\
\addlinespace
20 & raw & $0.035\ [0.029,0.038]$ & $1.00\ [0.99,1.01]$ & $0\ [0,1]$ & $1.30\ [1.14,1.42]$ & $11\ [9,27]$ & $224\ [222,264]$ \\
 & layer & $0.015\ [0.015,0.016]$ & $1.00\ [1.00,1.00]$ & $1\ [0,3]$ & $1.02\ [1.01,1.02]$ & $862\ [802,914]$ & $131\ [103,156]$ \\
 & \emph{uncorrected} & $0.067$ & $1.04$ & \multicolumn{4}{c}{(exact draws from the false factorization)} \\
\addlinespace
4 & raw & $0.025\ [0.023,0.040]$ & $1.00\ [1.00,1.01]$ & $0\ [0,0]$ & $1.09\ [1.05,1.16]$ & $48\ [27,65]$ & $172\ [164,206]$ \\
 & layer & $0.018\ [0.015,0.021]$ & $1.00\ [0.99,1.00]$ & $0\ [0,0]$ & $1.01\ [1.01,1.02]$ & $309\ [246,328]$ & $50\ [49,52]$ \\
 & \emph{uncorrected} & $0.219$ & $1.07$ & \multicolumn{4}{c}{(exact draws from the false factorization)} \\
\addlinespace
2.5 & raw & $0.079\ [0.073,0.080]$ & $0.99\ [0.98,1.00]$ & $1\ [0,4]$ & $1.21\ [1.21,1.29]$ & $15\ [11,15]$ & $182\ [161,237]$ \\
 & layer & $0.019\ [0.017,0.029]$ & $1.00\ [0.99,1.01]$ & $0\ [0,0]$ & $1.02\ [1.01,1.03]$ & $171\ [167,195]$ & $47\ [47,89]$ \\
 & \emph{uncorrected} & $0.277$ & $1.08$ & \multicolumn{4}{c}{(exact draws from the false factorization)} \\
\bottomrule
\end{tabular}
}
\end{table}

\end{document}